\documentclass[journal,10pt]{IEEEtran}
\usepackage{stmaryrd}
\IEEEoverridecommandlockouts
\usepackage{amsmath}
\usepackage{graphicx}
\usepackage{amssymb} 
\usepackage{amsmath,amsthm} 
\usepackage{float}
\usepackage{enumerate}
\usepackage{epstopdf}
\usepackage{dblfloatfix}
\usepackage{caption}
\usepackage{amsfonts}
\usepackage{fancyhdr}
\usepackage{bbm}
\usepackage{cases}
\usepackage{cite}
\usepackage{breqn}
\usepackage{multirow}
\usepackage{mathrsfs}
\usepackage{algpascal}
\usepackage{algc}
\usepackage{algorithmicx}
\usepackage[ruled]{algorithm}
\usepackage{algpseudocode}
\usepackage{graphics}
\usepackage{epsfig}
\usepackage{mathrsfs} 
\usepackage{dsfont}
\usepackage{color}
\usepackage{hyperref}

\newcommand{\tocaption}{%
\setlength{\abovecaptionskip}{0pt}%
\setlength{\belowcaptionskip}{10pt}%
\caption}

\floatname{algorithm}{Algorithm}

\makeatletter
\newcommand{\multiline}[1]{%
  \begin{tabularx}{\dimexpr\linewidth-\ALG@thistlm}[t]{@{}X@{}}
    #1
  \end{tabularx}
}
\makeatother

\algdef{SE}[DOWHILE]{Do}{doWhile}{\algorithmicdo}[1]{\algorithmicwhile\ #1}%

\newtheoremstyle{theorem}
  {\topsep}
  {\topsep}
  {\itshape}
  {}
  {\itshape}
  {:}
  {.5em}
  {\thmname{#1}\thmnumber{ #2}\thmnote{ (#3)}}
\theoremstyle{theorem}
\newtheorem{theorem}{Theorem}

\newtheorem{lemma}{Lemma}

\ifCLASSINFOpdf
 \else
\fi

\begin{document}
\title{Coverage and Rate Performance Analysis of Multi-RIS-Assisted Dual-Hop mmWave Networks}

\author{Yuwen Cao\textsuperscript{},~\IEEEmembership{Member, IEEE},
Xiaowen Wu\textsuperscript{},~\IEEEmembership{}Jiguang He\textsuperscript{},~\IEEEmembership{Senior Member, IEEE},
Tomoaki Ohtsuki\textsuperscript{},~\IEEEmembership{Senior Member, IEEE}, and Tony Q. S. Quek\textsuperscript{},~\IEEEmembership{Fellow, IEEE}\vspace{-0.48cm}

\thanks{
The work of Y. Cao was supported in part by the National Natural Science Foundation of China under Grant 62301143, in part by Shanghai Sailing Program under Grant 23YF1400800, and also in part by the Fundamental Research Funds for the Central Universities under Grant 2232024D-38.
The work of J. He was supported by Guangdong Research Team for Communication and Sensing Integrated with Intelligent Computing (Project No. 2024KCXTD047).
The work of T. Q. S. Quek was supported by the National Research Foundation, Singapore and Infocomm Media Development Authority under its Future Communications Research \& Development Programme.}
%
\thanks{This paper was presented in part at IEEE VTC 2024-Fall \cite{10757887}.
Y. Cao and X. Wu are with the College of Information Science and Technology, Donghua University, Shanghai, China (e-mail:ywcao@dhu.edu.cn, 18370398529@163.com).  
J. He is with the School of Computing and Information Technology, Great
Bay University, Dongguan 523000, China, and Great Bay Institute for Advanced Study (GBIAS), Dongguan 523000, China (e-mail:jiguang.he@gbu.edu.cn). T. Ohtsuki is with the Department of Information and Computer Science, Keio University, Yokohama, Japan (e-mail:ohtsuki@ics.keio.ac.jp). T. Q. S. Quek is with the Information Systems Technology and Design Pillar, Singapore University of Technology and Design, Singapore (e-mail:tonyquek@sutd.edu.sg).}
}

{}

\maketitle

\begin{abstract}
Millimeter-wave (mmWave) communication, which operates at high frequencies, has gained extensive research interest due to its significantly wide spectrum and short wavelengths.
However, mmWave communication suffers from the notable drawbacks as follows: i) The mmWave signals are sensitive to the blockage, which is caused by the weak diffraction ability of mmWave propagation; ii) Even though the introduction of reconfigurable intelligent surfaces (RISs) can overcome the performance degradation caused by serve path loss, the location of users and RISs as well as their densities incur a significant impact on the coverage and rate performance;
iii) 
When the RISs' density is very high, i.e., the network becomes extremely dense, a user sees several line-of-sight RISs and thus experiences significant interference, which degrades the system performance.
Motivated by the challenges above, we first analyze distributed multi-RIS-aided mmWave communication system over Nakagami-$m$ fading from the stochastic geometry perspective. To be specific, we analyze the end-to-end (E2E) signal-to-interference-plus-noise-ratio (SINR) coverage and rate performance of the system. To improve the system performance in terms of the E2E SINR coverage probability and rate, we study the optimization of the phase-shifting control of the distributed RISs and optimize the E2E SINR coverage particularly when deploying a large number of reflecting elements in RISs. To facilitate the study, we optimize the dynamic association criterion between the RIS and destination.
Furthermore, we optimize the multi-RIS-user association based on the physical distances between the RISs and destination by exploiting the maximum-ratio transmission.
Numerical and simulation results indicate that the deployment of distributed RISs can significantly improve the E2E SINR coverage probability and achievable rate of the system compared to the selected benchmarks.
\end{abstract}


\section{Introduction}
With the increasingly tight spectrum resources, millimeter-wave (mmWave) technology has become the core technology of the next generation communication network thanks to its rich spectrum resources and limited interference, especially its large-scale multi-input multi-output (MIMO) capability with high spatial resolution to facilitate the centralized power transmission of signals\cite{lodro2019mmwave}.

However, signals from mmWave communication systems experience significant challenges in propagation, including poor transmission through the atmosphere, limited penetration of buildings and obstacles, and substantial attenuation caused by foliage and precipitation \cite{niu2015survey}. To address these challenges, one effective approach is to deploy distributed reconfigurable intelligent surfaces (RISs) \cite{huang2019reconfigurable,8811733}. By controlling the amplitude and phase shift of each element of RISs, the incident electromagnetic wave can be altered, thereby improving the propagation environment. In addition, distributed RISs provide additional signal reception links to ensure the quality of service to users via its tremendous beamforming gain.
Moreover, the energy consumption of RF-chain-free RIS is several orders of magnitude lower than that of traditional fully-digital active antenna arrays \cite{8796365}.

Nevertheless, analyzing the performance of distributed RIS-assisted mmWave communication systems remains a challenge. In recent years, stochastic geometry (SG) has become the focus of research on distributed systems due to its ability to represent the spatial randomness of different types of wireless networks, and to extend the uncertainty of other factors including shadows, fading, etc. In addition, SG has been identified as a meaningful tool by providing a unified mathematical framework for describing communication systems\cite{win2009mathematical,5226957,8972478,9201540}. In \cite{6932503}, the authors leveraged the line-of-sight (LoS) probability function proposed in \cite{docomo20165g} to model the locations of LoS and non-line-of-sight (NLoS) base stations (BSs) as two independent inhomogeneous Poisson point process (PPP), and further analyzed the dense network situation.
In \cite{xu2024stochastic}, the authors leveraged the PPP to model the locations of users, blockage and RISs, and analyzed the impact of distributed RISs on the ergodic coverage probability and sum rate.
The authors in \cite{rebato2019stochastic} introduced an analytical framework which includes realistic channel models and antenna element radiation patterns.
In \cite{9018083}, the authors studied scheduling strategies to optimize information freshness in wireless networks by exploiting the synergy of traditional queuing theory and SG.

Based on the aforementioned discussions, this paper
investigates the deployment of distributed multi-RIS systems in mmWave communications using SG technology, focusing on improving signal-to-interference-plus-noise-ratio (SINR) coverage and rate performance through phase shift optimization. In the following subsections, we first review the relevant research work, followed by summarizing the contributions of this paper.

\subsection{Related Work}
In recent years, 
the analysis of distributed RIS-assisted communication systems has gained lots of research attention \cite{10413214,10468784,10158937,10236498,10437628,10035513,9935107,9856592,9894109}. More concretely,
references \cite{10413214,10468784,10158937} focused on analyzing the overall outage probability, average bit error rate (BER) and average channel capacity performance of RIS-assisted communication systems. 
These works modeled the path loss for RIS reflective links in free-space scenarios, omitting considerations of scattering, signaling overhead, and RIS element configuration.
In \cite{10236498}, the authors adopt the airborne RIS system of unmanned aerial vehicle (UAV) transmitted by non-orthogonal multiple access (NOMA). In addition, this work analyzed the system performance through the deduced interrupt probability and user traversal rate. 
Liu \textit{et al.} \cite{10437628} claimed that results of system evaluation for RIS are not accurate when the number of reflection elements is limited. A comprehensive solution using the H-function of multiple Foxes was proposed to solve this problem. 
In \cite{10035513}, the performance of RIS-assisted wireless communication system was studied by considering both direct and reflective paths under the assumptions of perfect channel state information acquisition and optimal phase-shift
configuration at RIS. 
Moreover, an adaptive RIS clustering system was proposed in \cite{9935107}, in which the RIS in the cluster can improve the overall performance of the system by reflecting signals under non-ideal conditions. In \cite{9856592}, the authors proposed a physics-based, parameterized end-to-end (E2E) tunable fading model for wireless channels incorporating RIS. By integrating the concepts of frequency selectivity and the amplitude-frequency response of RIS components, the model addressed a critical limitation: the inability of conventional channel models used for signal processing algorithm analysis to account for underlying physical principles. In the above cases, Gaussian noise was considered. In order to fill the gap of the impact of non-Gaussian noise on the RIS-assisted communication system, the authors studied the impact of pulse noise on the RIS-assisted communication system in \cite{9894109}. Notably, \cite{9856592} and \cite{9894109} introduced RIS-parametrized
wireless channels with adjustable fading, neglecting considerations of RIS design and deployment.

Note that there has a few works focused on geometry-based analysis of the distributed RIS-assisted large networks \cite{xu2024stochastic}, \cite{9606895,9224676,9110835,9174910}.
In \cite{9606895}, the authors studied the coverage probability and ergodic rate of the RIS-aided NOMA system via SG, and demonstrated the effectiveness of RIS-aided NOMA system by comparing the performance of RIS-aided NOMA with that of traditional orthogonal multiple access (OMA). 
Furthermore, reference \cite{9224676}  studied the coverage performance of RIS-aided large-scale mmWave networks using SG and derived the peak received power of RIS and the downlink signal-to-interference ratio coverage in closed forms. Likewise, the authors in \cite{9110835} analyzed the performance of large-scale mmWave networks supported by large intelligent surfaces. Besides, the authors verified the effectiveness of the designed networks via analyzing the average achievable rate, area spectral efficiency (ASE), and energy efficiency (EE), based on the assumption that there exists an optimal number of elements in large intelligent surfaces and density that maximizes ASE and EE. Similar to \cite{9110835}, the authors in \cite{9174910} focused on the SG-based analysis of RIS-enabled
communication systems. In addition, this work derived an upper bound of the RIS deployment efficiency. 
These aforementioned works consistently verified that RISs can theoretically provide unprecedented improvements in both the achievable rate and EE.

\subsection{Summary of Our Contributions}
It is noteworthy that, the RIS-user association
design is also coupled with \textit{phase-shifting control} at the distributed RISs, and the \textit{directional beamforming} at the BS, since directional beamforming at the BS may change the optimal phase-shifting control and the RIS-user associations, and vice versa. Thus, the \textit{dynamic RIS-user/BS
association} design for distributed RIS-assisted mmWave communications over Nagakami-$m$ fading is a new and non-trivial problem, which, however, has not been
studied in the literature to the best of our knowledge. Note also that distributed RIS-assisted mmWave network has been
recently studied in a handful of related works \cite{xu2024stochastic,10468784,10158937,10437628,10035513,9935107,9606895,9110835,9174910}.
However, these works ignore the RIS-user/BS association and thus do not investigate their optimal design along with other key system parameters such as BS directional beamforming or RIS phase-shifting control. Besides, how to further improve the coverage probability and
achievable rate in distributed RIS-assisted large networks through proper RIS phase-shifting control deserves further research.

Motivated by the abovementioned challenges, this paper studies the effective gain of the SINR coverage probability and achievable rate of dual-hop mmWave communication system with the assistance 
 of distributed RISs.
Specifically, based on the SG concept, we first develop a distributed-RIS-aided mmWave
communication system over Nakagami-$m$ fading. Afterwards, we analyze the SINR coverage probability and the achievable rate performance for the distributed RIS-assisted mmWave communications. Next, to improve the performance in terms of the SINR coverage probability and achievable rate, we jointly optimize the RIS phase shifter and the dynamic RIS-user/BS association criterion.
Notably, the specific contributions are summarized as follows:


\begin{itemize}
\item We first consider two cases where the typical user (TU) communicates directly with the BS or via RIS reflection links. In the case that the BS communicates with the user through the RIS, we determine the location of the associated RIS by introducing the association criterion, where the TU has the maximum E2E SINR. For further explanation, the probability density function (PDF) of the transmission distance between TU and its associated RIS is further obtained according to the derived association probability and LoS probability. The corresponding expected distance is obtained according to the expectation formula.
Combined with the constraints of RIS distribution, the specific position of the associated RIS is obtained as well.
\item We adopt the SINR coverage probability and achievable rate as performance metrics to measure the quality of the communication system model. The \textit{traversal coverage} of the system is derived by analyzing the SINR of the received signal and the given threshold. In addition, we obtain the expression of the achievable rate of the system by using Campbell's theorem.
\item{We provide the closed-form expressions for the PDF and
cumulative distribution function (CDF) of the E2E BS-RIS $n$-TU reflective channel, which are used to calculate the ergodic coverage probability and the ergodic rate.}
\item Numerical and simulation results reveal that the deployment of distributed RISs can significantly improve the SINR coverage probability and achievable rate of the system compared to the selected benchmarks. In particular, our approach enables the SINR
coverage probability performance improvement by 50.7$\%$ when properly deploying the distributed RISs with a density of $1.5e{-3} /\mathrm{m}^{2}$. 
\item Besides, we evaluate the SINR coverage probability and sum-rate performance via the Monte Carlo simulations, which turn out to be well-aligned to the theoretical results.
\end{itemize}

The remainder of this paper is organized as follows.
In Section~\ref{sec:sm}, the multi-RIS-aided dual-hop
mmWave communication system assumed in this paper is described. Afterwards,
an optimal phase-shifting control strategy and the directional beamforming gain are elaborated in Section~\ref{sec:phase}. 
The optimal phase-shifter of multi-RIS system as well as the dynamic RIS-user/BS association criteria are designed and proposed in  Section~\ref{sec:opsc}, respectively.
Furthermore, the analysis of the theoretical performance in terms of the E2E SINR coverage
probability and ergodic rate is provided in Section~\ref{sec:pa}. 
In Section~\ref{sec:simulations}, the performance of our proposed schemes is evaluated.
Finally, the conclusion of this paper is given in Section~\ref{sec:conclusion}.

{\it Mathematical Notations}: Bold lower case and upper case letters are used to represent vectors and matrices, respectively.
$\left( {\cdot} \right)^{T}$, $\left( {\cdot} \right)^{H}$, $\rm{Tr}\left[ {\cdot} \right]$, $\left\| {\cdot} \right\|_{2}$  and $\left\| {\cdot} \right\|_{\rm{F}}$ denote transpose, Hermitian, the trace of a square matrix, $l_{2}$-norm and Frobenius norm operations, respectively.
The $i$-th row and $j$-th column entry
of matrix $\mathbf{C}$ is $[\mathbf{C}]_{i,j}$, whereas the $j$-th column of $\mathbf{C}$ is $[\mathbf{C}]_{:,j}$.
$\mathbb{E}[{\cdot}]$ means the expectation of the argument, whereas $\mathrm{Var}[{\cdot}]$ denotes the variance of the argument.
$\textrm{diag}(\mathbf{x})$ corresponds to a square diagonal matrix with entries of $\mathbf{x}$ on its diagonal, 
whereas $\textrm{bdiag}(\mathbf{X}_1,\ldots,\mathbf{X}_{N})$ creates a block diagonal matrix with entries of matrices $\mathbf{X}_{1},\ldots,\mathbf{X}_{N}$ on its diagonal.
$\textrm{vec}(\mathbf{X})$ represents the vectorization of $\mathbf{X}$ by stacking the columns of the matrix $\mathbf{X}$ on top of one another.
$\tiny
\left( {\begin{array}{*{20}{c}}
{\cdot}\\
{\cdot}
\end{array}} \right)$ denotes the binomial coefficient.
Besides, $\mathbf{x} \circ \mathbf{y}$ and $\mathbf{x} \varotimes \mathbf{y}$ denote the Khatri-Rao product and Kronecker product between two complex vectors $\mathbf{x}$ and $\mathbf{y}$, respectively. $\mathbf{I}_{N}$ is used to denote an identity matrix of size $N\times N$.

\begin{figure}[tp]
    \begin{center}
    \includegraphics[width=0.35\textwidth]{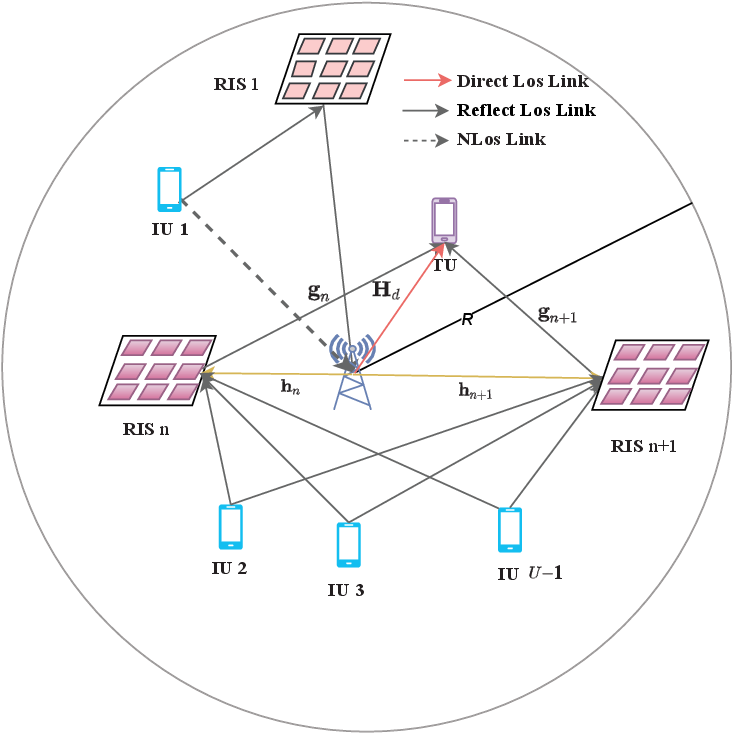}
    \end{center}
    \caption{An overview of the proposed distributed multi-RIS-assisted mmWave system. Herein, IU refers to the interfering user.} \label{fig:2}
\end{figure}

\section{System Model}
\label{sec:sm}
%
We consider a multi-RIS-aided dual-hop mmWave wireless communication system.
\footnote{In this paper, dual hop refers strictly to BS-RIS $n$ and RIS $n$-UE links, $n\in\{1,\ldots,N\}$, to maintain tractability.} We focus on downlink coverage and rate experienced by TU. In this model, we adopt the concept of the unit virtual radius, that is, the equivalent circle radius of the region, denoted as $R$.
In this paper, we assume that the location information of $U$ users is known in prior at the BS, and each user is equipped with $N_{u}$ antennas with $u\in\{1,\ldots,U\}$ and the BS is equipped with $N_{b}$ antennas. The locations of $U$ users are randomly distributed by following PPP, within the circle of specified radius $R$ of the BS, whereas the location distribution of RISs is modeled by a homogeneous PPP expressed as $\Phi_{R}=\{\mathbf{r}_{n}\}\in\mathbb{R}^{2}$, $n\in\{1,\ldots,N\}$, with a density of $\lambda_{R}$.\footnote{In Section \ref{sec:simulations}, we model the locations of RISs using the Poisson cluster process (PCP)\cite{10064007,6524460}, to reflect real-world clustering near BS/users, followed by comparing its coverage performance with that of the PPP model.} Herein, $\mathbf{r}_n$, $\Phi_R$, and $L_n$ represent the position of the $n$-th RIS, the set of all RIS coordinate points, and the number of passive reflecting elements of RIS $n$, respectively.

Note that Fig. \ref{fig:2} illustrates the distributed multi-RIS-assisted mmWave communication system, in which the baseband signal received at the TU can be mathematically expressed as
\begin{equation}
\label{eq:001}
\mathbf{Y}=\sqrt{P}x\left(\sum_{n=1}^N \sum_{l=1}^{L_{n}} \mathbf{g}_{n}  \Theta_{nl} \mathbf{h}_{n}^{T} +\mathbf{H}_d\right)+\mathbf{Z},
\end{equation}
where $P$ refers to the transmit power in dBm at the BS. Moreover, $\mathbf{h}_{n}\in\mathbb{C}^{N_{b}}$ and $\mathbf{g}_{n}\in\mathbb{C}^{N_{u}}$ denote the complex channel vectors from the BS to the RIS $n$, and from the RIS $n$ to the TU, respectively.
$\mathbf{H}_d\in\mathbb{C}^{N_{u}\times N_{b}}$ is the complex channel matrix from the BS to the TU. In addition, $x$ indicates the transmit signal at the BS with unit power, i.e., $\mathbb{E}[|x|^{2}] = 1$.
$\Theta_{nl}=\exp(j\theta_{nl})$ denotes the $l$-th reflecting element response of RIS $n$ with $\theta_{nl}\in[0,2\pi]$ being the phase shift coefficient.
$\mathbf{Z}\in\mathbb{C}^{N_{u}\times N_{b}}$ represents the additive white Gaussian noise matrix with each entry having zero mean and variance $\sigma^2$.

Note that the terms $h_{n,l}=|h_{n,l}|\exp(j\phi_{h_{n,l}})$, $g_{n,l}=|g_{n,l}|\exp(j\phi_{g_{n,l}})$, and $h_{d,u,i}=|h_{d,u,i}|\exp(j\phi_{h_{d,u,i}})$ introduced in (\ref{eq:001}) are assumed to be independent random variables (RVs) whose envelopes $|h_{n,l}|,|g_{n,l}|$, and $|h_{d,u,i}|$ follow Nakagami-$m$ distribution with parameters $(m_{h_{n,l}}, \Omega _{h_{n,l}}) $, $( m_{g_{n,l}}, \Omega _{g_{n,l}}) $, and $( m_{h_{d,u,i}}, \Omega _{h_{d,u,i}}) $, respectively. Herein, $u\in\{1,\dots,N_{u}\}$, $i\in\{1,\ldots,N_{b}\}$. The parameters $\Omega_{\ell}$ for $\ell\in\{h_{n,l},g_{n,l},h_{d,u,i}\}$, are 
given as $\Omega_\ell=\zeta \xi_\ell^{-\alpha}$ where $\zeta=\left(\frac{\lambda}{4\pi}\right)^{2}$ is the far-field path loss factor at a reference distance of l m, with $\lambda = \frac{c_0}{f_c}$ being the carrier wavelength, and $\alpha\geq2$ denoting the path loss exponent.
\footnote{
Note that the Nakagami-$m$ fading model assumes the power $|h_{n,l}|^2(|g_{n,l}|^2 ~\text{or}~ |h_{d,u,i}|^2)$ follows a Gamma distribution with mean $\Omega_{\ell} = \zeta \xi_\ell^{-\alpha}$, where $\zeta$ is the reference far-field path loss and $\xi_\ell$ is the distance. This indicates that the large-scale path loss is integrated into the channel power distribution in our fading settings.
}
Herein, $c_0$ means the speed of light, and $f_{c}$ is the centre frequency in Hz.
Finally, the phases $\phi_{h_{n,l}},\phi_{g_{n,l}}$, and $\phi_{h_{d,u,i}}$ are assumed to be independent from each other, as well as independent from the envelopes $\left|h_{n,l}\right|, \left|g_{n,l}\right|$ and $\left|h_{d,u,i}\right|$ \cite{9406837}.

In the following, we study the path loss in LoS and NLoS cases, respectively.
In our system model, the path loss with respect to (w.r.t) the NLoS link is integrated in the spread parameter of the Nakagami-$m$ distribution. 
To be specific, the path loss w.r.t the BS-TU link can be expressed as $\Omega_{\textrm{SD}}(\xi) = \Omega_{h_{d,u,i}} \cdot G_{\text{S}} \cdot G_{\text{D}}$ with $G_{\text{S}}(G_{\text{D}})$ being the antenna gain at the BS(TU), and $\xi$ denoting the spatial distance between the source and the destination in two-dimensional Cartesian coordinate systems; the path loss w.r.t. the BS-RIS $n$ link can be expressed as $\Omega_{\textrm{SRn}}(\xi) = \Omega_{h_{n,l}} \cdot G_{\text{S}} \cdot G_{\text{R{n}}}$ with $G_{\text{R{n}}}$ being the antenna gain at the $n$-th RIS; the path loss w.r.t. the RIS $n$-TU link can be expressed as $\Omega_{\textrm{RnD}}(\xi) = \Omega_{g_{n,l}} \cdot G_{\text{R{n}}} \cdot G_{\text{D}}$.
\footnote{
Multiple-hop communication with multi-RIS reflections will incur severe multi-path fading, as the cumulative pathloss with respect to the BS-RIS ($n$-1)-RIS $n$-RIS ($n$+1)-UE link will become very big, thus leading to destructive effect of the multi-path fading. Therefore, we only consider dual-hop concept in multi-RIS-aided networks. Performance analysis of multiple-hop communication with multi-RIS reflections taking into account the impact of both the complexity, multi-path fading, as well as the practical gain will be studied in our future research.
}

To overcome the \textit{destructive effect} of multi-path fading, the phase-shifts of the strongest RIS are reconfigured such that the received signals of the dual-hop communications are constructively added to achieve the largest SINR at TU. Mathematically, the optimal phase-shift configuration of the $l$-th reflecting element of the $n$-th RIS can be
expressed as
\begin{align}
\Theta_{nl}^{\star}=\arg\max_{\theta_{nl}\in\boldsymbol{\Theta}} \gamma(\Theta_{nl}),\quad\forall l,\quad\forall n,
\end{align}
where $\gamma(\Theta_{nl})$ represents the received SINR at TU as a function of the phase-shifts of the desired RIS. Note that $\gamma(\Theta_{nl})$ is RV and the statistical characterization of the E2E SINR will be analyzed and provided in Secs. III and V. 
Note also that the proposed optimal phase-shifting control approach and the association criterion between the BS, RISs, and TU will be detailed in the following Secs. III and IV, respectively. It should
be noted that, the optimal deployment strategy of the RIS,
closing to the BS or the TU, yields enhanced link gains,
as demonstrated in \cite{9693963}.
When taking into account the case of deploying all $N$ RISs, we observe the following insightful results. 
\begin{itemize}
    \item When increasing the RIS density, $i.e., \lambda_{R}$, the distances between the user and RIS become smaller, and the user becomes more likely to be associated with a LoS RIS. When the density is very high, i.e., $N\to \infty$, however, a user sees several LoS RISs and thus experiences significant interference.
    \item  Besides, when employing multiple RISs, the design of association criteria using the maximum-ratio transmission (MRT) mechanism is critical and desired to be explored.
\end{itemize}

\section{Phase-Shifting Control and E2E Channel Directivity Gain}
\label{sec:phase}
\subsection{Optimal Phase-Shifting Control}

Note that the objective of this paper is to design the phase control matrix of the involved multi-RIS to maximize the E2E SINR at the TU, while satisfying the unit-modulus constraint. To this end, we attempt to maximize the power of the E2E channel matrix 
$\sum_{l=1}^{L_{n}} \mathbf{g}_{n}  \Theta_{nl} \mathbf{h}_{n}^{T} \in \mathbb{C}^{N_{u}\times N_{b}}$
as a function of $\Theta_{nl}$, i.e., $\|\sum_{l=1}^{L_{n}} \mathbf{g}_{n}  \Theta_{nl} \mathbf{h}_{n}^{T}\|^{2}_{\mathrm{F}}$. 
In this case, the optimal design of the RIS reflecting element response matrix denoted by $\boldsymbol{\Theta}$ is equivalent to solving the problem of
\begin{align}
\label{eq:0003}
\boldsymbol{\Theta}^{\star} = \arg\max_{\boldsymbol{\Theta}} \;\Big\|\sum_{l=1}^{L_{n}} \mathbf{g}_{n}  \Theta_{nl} \mathbf{h}_{n}^{T}\Big\|_{\mathrm{F}}^{2},
\end{align}
where the term $\|\sum_{l=1}^{L_{n}} \mathbf{g}_{n}  \Theta_{nl} \mathbf{h}_{n}^{T}\|_{\mathrm{F}}^{2}$ can be further expressed as
\begin{align}
\label{eq:010}
\begin{split}
\Big\|\sum\limits_{l=1}^{L_{n}} \mathbf{g}_{n}  \Theta_{nl} \mathbf{h}_{n}^{T}\Big\|_{\mathrm{F}}^{2} & = \Big\|\mathrm{vec}\Big(\sum\limits_{l=1}^{L_{n}} \mathbf{g}_{n}  \Theta_{nl} \mathbf{h}_{n}^{T}\Big)\Big\|_{{2}}^{2} \\
&= \|\mathrm{vec}\left(\mathbf{G} \boldsymbol{\Theta} \mathbf{H}^{T}\right)\|_{{2}}^{2} \\
&\overset{(\text{I})}{=}\|\left(\mathbf{H} \varotimes \mathbf{G}^{} \right) \boldsymbol{\omega} \|_{{2}}^{2}\\
&= \|\left(\mathbf{H} \circ \mathbf{G}^{} \right) \underline{\boldsymbol{\omega}}\|_{{2}}^{2},
\end{split}
\end{align}
where $\mathbf{H}$ and $\mathbf{G}$ correspond to the channel matrix of the BS-RIS $n$ link, and the channel matrix of the RIS $n$-TU link, respectively.
Besides, $\boldsymbol{\omega} = \mathrm{vec}(\boldsymbol{\Theta})$ and $\underline{\boldsymbol{\omega}} = \mathrm{diag}(\boldsymbol{\Theta})$. Notably, the step (I) in (\ref{eq:010}) is based on the property that $\mathrm{vec}\left(\mathbf{A}\mathbf{B}\mathbf{C}\right) = \left(\mathbf{C}^{T} \varotimes \mathbf{A} \right)\mathrm{vec}\left(\mathbf{B}\right)$ \cite{9354904}.

Based on the above observations, we observe that the optimal phase-shift vector $\underline{\boldsymbol{\omega}}^{\star}$ is obtained by means of solving the problem of
\begin{align}
\label{eq:0005}
\begin{split}
\underline{\boldsymbol{\omega}}^{\star}& =\arg\max_{\boldsymbol{\omega}}  \|\left( \mathbf{H} \circ \mathbf{G}^{} \right) \underline{\boldsymbol{\omega}}\|_{{2}}^{2}\\
%
&=\arg\max_{\boldsymbol{\omega}} \|\mathbf{E} \,\underline{\boldsymbol{\omega}}\|_{{2}}^{2},
\end{split}
\end{align}
where the term $\mathbf{E} = \mathbf{H} \circ \mathbf{G}^{}$.
Afterward, we perform the singular value decomposition (SVD) operation on $\mathbf{E}$, i.e., $\mathbf{E} = \mathbf{U}\mathbf{D}\mathbf{V}^{H}$, where $\mathbf{D}$ is a diagonal matrix with singular values on the diagonal as a descending order and 
$\mathbf{V}\mathbf{V}^{H}$ = $\mathbf{V}^{H}\mathbf{V}$ = $\mathbf{I}_{L_{R,T}}$, as well as $\mathbf{U}\mathbf{U}^{H}$ = $\mathbf{U}^{H}\mathbf{U}$ = $\mathbf{I}_{L_{B,R}}$, with $L_{R,T} = \min(N_{u}, L_{n})$ and $L_{B,R} = \min(N_b, L_{n})$ being the rank of matrices $\mathbf{G}$ and $\mathbf{H}$, respectively. We observe that the optimal phase shifts of RIS, denoted by $\underline{\boldsymbol{\omega}}^{\star}$, is achieved by taking the first column of $\mathbf{V}$ (i.e., $[\mathbf{V}]_{:,1}$), followed by projecting $[\mathbf{V}]_{:,1}$ to the unit-modulus vector space. In other words, we conduct $\underline{\boldsymbol{\omega}}^{\star} = \mathrm{exp}(j\mathrm{Phase}([\mathbf{V}]_{:,1}))$, where $\mathrm{Phase}(\cdot)$ means the element-wise operation of extracting the phases of the argument.

By inspection, we observe that: i) to accurately reconfigure the phase-shifts of RISs, lots of novel phase-shifting control strategies have been developed in  \cite{9558795},\cite{9205879},\cite{he2021channel}. In particular, the destination in \cite{9205879} needs to process $(N L_{n}+1)$ pilots to estimate the ideal phase-shift configuration exhaustively. In \cite{9558795}, the destination needs to process a total number of $(L_{n}+1)$ pilots to estimate the phase-shift configuration. In our work, the destination just needs to process $L_{n}$ pilots. ii) Prior works design the discrete phase-shift configuration method. However, the number of the RIS reflecting elements usually is limited and constrained by the phase-shift resolution. Such strategy is invalid for multiple RIS-aided networks installing a large number of
reflecting elements, and can not guarantee maximum achievable rate for multiple RIS-aided networks. iii) The proposed optimal RIS phase-shifting control approach is valid for both the multi-RIS-aided systems \cite{8796365},\cite{10437628}, and multi-hop multi-RIS-aided networks \cite{9558795}.

Note that performance analysis on RIS-assisted large networks has been studied in references \cite{xu2024stochastic}, \cite{10413214,10468784,10158937,10236498,10437628,10035513,9935107,9856592,9894109}, \cite{9224676,9110835,9174910}. Specifically, references \cite{10158937,10236498,10035513} investigate the performance of RIS-assisted communication models assuming perfect channel state information and optimal phase-shift configuration at both the RIS controller, BS, and users. \cite{xu2024stochastic}, \cite{9224676,9110835,9174910} focus on the geometry-based analysis of distributed RIS-assisted networks for improving either the achievable rate or EE, omitting considerations of intra-cell interference, RIS design, RIS deployment, and RIS-user association. In this paper, we investigate the deployment of distributed multi-RIS systems in mmWave networks using SG, focusing on improving the coverage and rate performance through optimizing the RIS phase shift and the dynamic RIS-user/BS association criterion. In addition, to address the computational complexity and accuracy trade-offs, we develop a sub-optimal RIS phase-shifting control method that approximates the optimal RIS phase-shift optimization solution.


\subsection{Directional Beamforming Gain}
In the following, we first analyze the small-scale fading and directional beam pattern considered in our model. Afterwards, we provide an approximated directivity gain of each E2E channel link under distinct cases of the link occurring probabilities, to ease the analysis of the characteristic of the SINR statistics
as well as the ergodic coverage probability in distributed RIS-assisted mmWave communications.

\textbf{Assumption 1:} (\textit{Directional Beamforming Gain}) The directional beamforming gain is approximated by the segmented model, i.e.,
\begin{align}
\rho_{t,r}=
\begin{cases}M_tM_r, \;\text{w.p.} \;\frac{\psi_t}{2\pi} \cdot \frac{\psi_r}{2\pi},\\M_tm_r, \;\text{w.p.} \;\frac{\psi_t}{2\pi} \cdot (1-\frac{\psi_r}{2\pi}),\\m_tM_r, \;\text{w.p.}\;  (1-\frac{\psi_t}{2\pi}) \cdot \frac{\psi_r}{2\pi},\\m_tm_r, \;\text{w.p.}\; (1-\frac{\psi_t}{2\pi})\cdot (1-\frac{\psi_r}{2\pi}),
\end{cases}
\end{align}
where the subscripts $t$ and $r$ account for the transmitter (source) and receiver (destination), respectively. $M_{t}\mathrm{~and~}M_{r}$ represent their main-lobe directivity gain, while $\psi_t\mathrm{~and~}\psi_r$ represent their corresponding main lobe beam widths at the source- and destination-sides, respectively. We adopt the isotropic antennas, and the main lobe directivity gains are equal to the number of antenna elements, i.e., $M_{t}=N_{b}$, $M_{r}=N_{u}$. Meanwhile, the side lobe gains are expressed as $m_t=1/\sin^2(\frac{3\pi}{2\sqrt{N_b}})$ and $m_{r}=1/\sin^{2}(\frac{3\pi}{2\sqrt{N_{u}}})$.

\textbf{Assumption 2:} \textit{(Small-Scale Fading and Beamforming)} We assume independent Nakagami-$m$ fading for each link. Different parameters
of Nakagami-$m$ fading $(m_{h_{n,l}}, \Omega _{h_{n,l}}) $, $( m_{g_{n,l}}, \Omega _{g_{n,l}}) $, and $( m_{h_{d,u,i}}, \Omega _{h_{d,u,i}}) $ are assumed for both LoS and NLoS links. Recall that $h_{n,l}$ is the small-scale fading term from the BS to the $l$-th reflecting element of RIS $n$. Then, $|h_{n,l}|^2$ is a normalized Gamma RV. In addition, we employ beamforming to improve the SINR coverage and rate performance of the communication system. We assume in our system that the beams involved at the source and the destination are well aligned, whereas the beams between the source/destination and the unselected (i.e., the non-serving) intermediates (such as the NLoS RISs) are misaligned, meaning that the corresponding angle of arrival (AoA)
and angle of departure (AoD) are RVs with uniform distribution. In this case, such kind of interference from the IU $k$-BS link and/or IU $k$-NLoS RIS-BS link,  $k=2,\ldots,U$, cannot be ignored.

Thus, the SINR of the direct links and the RIS reflection links can be respectively defined as follows:
\begin{equation}
\label{eq:009}
\gamma_D:=\frac1{\sigma^2+\text{Interference}}|h_{d,u,i}|^2\cdot M_rM_t \cdot P\cdot \Omega_{\text{SD}}(\xi),\end{equation}
\begin{equation}
\label{eq:0100}
\begin{split}
\gamma_I:= &\frac1{\sigma^2+\text{Interference}}\left|\sum_{l=1}^{L_{n}} h_{n,l} \cdot [\underline{\boldsymbol{\omega}}^{\star}]_{l} \cdot g_{n,l} \right|^2  M_r M_t  P\\
&\cdot \Omega_{\text{S} \text{Rn}}(\xi) \cdot \Omega_{\text{R{n}D}}(\xi),
\end{split}
\end{equation}
with $[\underline{\boldsymbol{\omega}}^{\star}]_{l}$ being the $l$-th reflecting element response of RIS $n$ associated with the optimal phase shifts obtained by (\ref{eq:0005}), $n\in\{1,\ldots,N\}$.
For interference, we consider an extreme scenario where $U-1$ IUs interfere TU at the same time in the distributed multi-RIS-assisted
mmWave system. In this case, 
we formalize the interference signal from the IU in the form of \footnote{
Note that in our system the BS serves the TU in the downlink, while at the same time several users, i.e., the IU $k$, $k =2,\ldots,U$, transmit signal to BS in the uplink. It is noted that in Eq. (\ref{eq:011}), taking the $U-1$ IUs into accounts, in practice,  is an extreme case, although such scenario occurs with low probability. As such, the corresponding lower bound SINR of direct links and RIS reflection links as well as the rate are derived in this paper.}
\begin{equation}
\label{eq:011}
\text{Interference}:=\sum_{k=2}^U |h_k |^2\cdot \rho_{t,r} \cdot P\cdot \Omega_{\text{SI}}(\xi_{k}),
\end{equation}
where $\Omega_{\text{SI}}(\xi_{k})$ denotes the path loss of the $k$-th IU, as a function of the spatial distance $\xi_{k}$ from the source (i.e., BS or RIS) to the IU $k$. We note that the AoA and AoD follow a uniform distribution $\phi \sim \mathcal{U}[0,2\pi]$.
For the interference part of the channel $h_k$, since we do not know whether the interference signal reaches the TU through a direct link or a reflective link, $h_k$ is determined according to the following probabilities, i.e.,
\begin{equation}
\label{eq:012}
h_k=\begin{cases}h_{\text{d}},\;\text{w.p.}\;\mathrm{Pr}_{\mathrm{LoS}}(\xi_{k}),\\h_{\text{r}}, \;\text{w.p.} \; 1-\mathrm{Pr}_{\mathrm{LoS}}(\xi_{k}),\end{cases}
\end{equation}
with $h_{\text{d}}$ and $h_{\text{r}}$ being the LoS and NLoS link channels mentioned in \textbf{Assumption 2}, respectively. $\mathrm{Pr}_{\mathrm{LoS}}(\xi_{k})$ is the probability of LoS link as defined in (\ref{eq:09}). 
In this sense, the lower bound SINR received at TU after dual-hop communications yields
$\gamma= \gamma_{D} + \gamma_{I}$.

\section{Optimal Phase-Shifting Control of Distributed RISs}
\label{sec:opsc}
In this section, we first properly select a set of strong RISs based on the physical distances of BS-RIS-$n$, and the physical distances of RIS-$n$-TU, with $n=1,\ldots,N$. Afterwards, we reconfigure the phase-shifting control of the selected RISs to improve the successful reflective ratio of the dual-hop mmWave communications. Subsequently, we derive the directional beamforming by using the MRT, as well as analyze the directional beam patterns of the multiple RISs assisted dual-hop mmWave communications. We note that the directional beamforming can be implemented as the MRT based on the effective BS-RIS $n$-TU channel. Finally, we analyze the scalability of the proposed RIS phase-shifting control strategies. 

\subsection{Phase-Shifting Control of Distributed RISs}
\label{suc:opti}

To assist the dual-hop mmWave communications associated with a maximum received signal power at the TU, as well as to maximize the E2E SINR at the TU, a set of RISs are properly selected in our system. To this end, it is critical to properly reconfigure the phases of the involved RISs with a given selected RIS set, which are picked out in advance based on the
multi-RIS-user/BS association criteria introduced in the following Sec. IV-B. For convenience, we denote the sets of RISs by $\mathcal{K} = \{1,\ldots,K\}$ assuming that $K$ strong RISs are selected, satisfying $K \leq N$, for each configuration of the phase shift of $K$ RISs. Besides, we let $a_{k}$ be the index of the $k$-th selected RIS with $k\in\mathcal{K}$. It is noted that,
the RISs that are not selected can be
switched off and then regarded as \textit{random scatters} in the system,
which may result in additional signal paths from the BS to the
user. Nevertheless, these randomly scattered paths generally have
much lower strength as compared to the constructed LoS link thanks to the inter-RIS directional beamforming gain, particularly for the case of practically large $L_{k}$ with $k\in\mathcal{K}$.

Recall that the main objective of this paper is to devise the phase control matrix of the involved RISs to maximize the E2E SINR received at the TU, while satisfying the unit-modulus constraint. Based on (\ref{eq:001}) and with a given selected RIS set $\mathcal{K}$, the BS-RIS $n$-TU multi-reflective channel can be expressed as ${\mathbf{G}}_{\text{E2E}} = \sum_{n=a_{1}}^{a_{K}} \sum_{l=1}^{L_{n}} \mathbf{g}_{n}  \Theta_{nl} \mathbf{h}_{n}^{T} = \sum_{n=a_{1}}^{a_{K}}  \mathbf{G}_{n}\boldsymbol{\Theta}_{n}\mathbf{H}^{T}_{n} = \mathbf{G}_{\text{E2E}, a_{1}} + \ldots + \mathbf{G}_{\text{E2E},a_{K}}$, where $\mathbf{G}_{\text{E2E},a_{k}}$ refers to the E2E BS-RIS $a_k$-TU channel matrix. Next, we let $\hat{\bar{\mathbf{G}}} =[\mathbf{G}_{a_{1}}, \ldots,\mathbf{G}_{a_{K}}]$, and let $\hat{\bar{\mathbf{H}}} = [\mathbf{H}^{}_{a_{1}}, \ldots,\mathbf{H}^{}_{a_{K}}]$. Hence, we can arrive at that ${\mathbf{G}}_{\text{E2E}} = \sum_{n=a_{1}}^{a_{K}} \sum_{l=1}^{L_{n}} \mathbf{g}_{n}  \Theta_{nl} \mathbf{h}_{n}^{T} = \hat{\bar{\mathbf{G}}}  \hat{\bar{\boldsymbol{\Theta}}}\hat{\bar{\mathbf{H}}}^{T}$, where $\hat{\bar{\boldsymbol{\Theta}}}$ is defined as 
\begin{align}
\label{eq:0035}
\hat{\bar{\boldsymbol{\Theta}}} = \text{bdiag}(\boldsymbol{\Theta}_{a_{1}},\ldots, \boldsymbol{\Theta}_{a_{K}}).
\end{align}

On the other hand, by following (\ref{eq:0003}) we observe that maximizing the E2E SINR received at the TU amounts to 
maximizing the power of the BS-RIS $n$-TU multi-reflective channel matrix $\mathbf{G}_{\text{E2E}} = \hat{\bar{\mathbf{G}}}\hat{\bar{\boldsymbol{\Theta}}}\hat{\bar{\mathbf{H}}}^{T} \in \mathbb{C}^{KN_{u}\times KN_{b}}$ as a function of the $\boldsymbol{\Theta}_{a_k}$, i.e., $\left\|\mathbf{G}_{\text{E2E}}\right\|^{2}_{\mathrm{F}}$. In this case, the optimal design of $\boldsymbol{\Theta}_{a_k}$ w.r.t. the RIS $a_{k}$, $k\in\mathcal{K}$, is equivalent to solving the optimization problem of
\begin{align}
\begin{split}
\label{eq:0035}
\boldsymbol{\Theta}_{a_{k}}^{\star} &= \arg\max_{\hat{\bar{\boldsymbol{\Theta}}}, k \in \mathcal{K}} \; \| \hat{\bar{\mathbf{G}}}  \text{bdiag}(\boldsymbol{\Theta}_{a_{1}},\ldots, \boldsymbol{\Theta}_{a_{K}}) \hat{\bar{\mathbf{H}}}^{T} \|_{\mathrm{F}}^{2}\\
&= \arg\max_{\hat{\bar{\boldsymbol{\Theta}}}, k \in \mathcal{K}} \; \left\|\mathrm{vec}\left(\hat{\bar{\mathbf{G}}} \hat{\bar{\boldsymbol{\Theta}}} \hat{\bar{\mathbf{H}}}^{T}\right)\right\|_{{2}}^{2} \\
&\overset{(\text{III})}{=} \arg\max_{\hat{\bar{\boldsymbol{\Theta}}}, k \in \mathcal{K}} \; \left\|\left(\hat{\bar{\mathbf{H}}} \varotimes \hat{\bar{\mathbf{G}}}^{} \right) \hat{\bar{\boldsymbol{\omega}}} \right\|_{{2}}^{2}\\
&= \arg\max_{\hat{\bar{\boldsymbol{\Theta}}}, k \in \mathcal{K}} \; \left\|\left(\hat{\bar{\mathbf{H}}} \circ \hat{\bar{\mathbf{G}}}^{} \right) \underline{\hat{\bar{\boldsymbol{\omega}}}}\right\|_{{2}}^{2},
\end{split}
\end{align}
where $\hat{\bar{\boldsymbol{\omega}}} = \mathrm{vec}(\boldsymbol{\Theta}_{a_1}, \ldots, \boldsymbol{\Theta}_{a_{K}})$ and $\hat{\bar{\underline{\boldsymbol{\omega}}}} = \mathrm{bdiag}(\boldsymbol{\Theta}_{a_1}, \ldots, \boldsymbol{\Theta}_{a_{K}})$.
Notably, the step (III) in (\ref{eq:0035}) is based on the same property as that in the formula (\ref{eq:010}). 

Based on the aforementioned analysis, we observe that the optimal phase-shift vector $\hat{\bar{\underline{\boldsymbol{\omega}}}}^{\star}$ is obtained by maximizing the power of the E2E BS-RIS $n$-TU channel, i.e., $\left( \hat{\bar{\mathbf{H}}} \circ \hat{\bar{\mathbf{G}}}^{} \right) \hat{\bar{\underline{\boldsymbol{\omega}}}}$, according to the following phases: (i) the SVD operation on $\hat{\bar{\mathbf{H}}} \circ \hat{\bar{\mathbf{G}}}^{}$, (ii) taking the first-column of the right singular matrix, and (iii) performing the element-wise operation by extracting the phases of the argument in (\ref{eq:0035}) accordingly.
We note that optimizing the RIS phase shifters jointly in (\ref{eq:0035}) can enable higher E2E SINR gain than optimizing the RIS shifters separately by using the prior phase shifting control methods \cite{he2021channel,10437628,9558795}.

\subsection{Sub-Optimal Online Phase-Shifting Control}
\label{suc:subo}

Solving the SVD-based phase-shift optimization problem can be numerically costly for big matrices involved in 
large-scale networks assisted by RISs that are equipped with a large
number of reflecting elements, i.e., $L_n$, or in dynamic environments. 
To tackle the computational cost and accuracy trade-offs,  we attempt to approximate the optimal RIS phase-shift optimization solution by transforming the original high-order SVD problem in (\ref{eq:0035}) into a lower-order SVD problem, thus mitigating the computation cost through reducing the storage and data movements in implementation.

To address the computational cost and accuracy trade-offs, as well as to enable a real-time implementation, we develop a sub-optimal approach. \footnote{
In this paper, we assume that perfect channel state information is available at the BS and RIS controllers. Similar assumption has also been made in references \cite{10468784,10158937,10437628,10035513,9935107,9606895,9110835,9174910}. The prior information of RIS phase shifts is unknown in this paper for improving the feasibility for real-time implementation. In addition, related research focusing on addressing the gap between ideal control and practical hardware limitations will be studied in our future work. 
} Specifically, we first recursively define $\boldsymbol{\varpi}_{k+1} = [\boldsymbol{\varpi}_k, ~\boldsymbol{\zeta}_{k+1}]$ with a random renormalization vector
$\boldsymbol{\zeta}_{k+1}$, $k = 1,\ldots,KL_{n}$, and $\boldsymbol{\varpi}_k = [\boldsymbol{\zeta}_{1},\ldots,\boldsymbol{\zeta}_{k}]$. By invoking the QB decomposition \cite{GolubMatrix}, we have $(\hat{\bar{\mathbf{H}}} \circ \hat{\bar{\mathbf{G}}}^{} )\boldsymbol{\varpi}_{k+1} = \hat{\bar{\mathbf{Q}}}_{k+1}\hat{\bar{\mathbf{B}}}_{k+1}$.
Once the index $\tau$ appears such that $\hat{\bar{\mathbf{B}}}_{\tau}$ has no zero entries while $\hat{\bar{\mathbf{B}}}_{\tau+1}$ does, the above iteration process terminates and outputs $\hat{\bar{\mathbf{Q}}}_{\tau}$. 
Then, let $\hat{\bar{\mathbf{B}}} = \hat{\bar{\mathbf{Q}}}_{\tau}^{H} (\hat{\bar{\mathbf{H}}} \circ \hat{\bar{\mathbf{G}}}^{} )$, and $\hat{\bar{\mathbf{C}}} = \hat{\bar{\mathbf{B}}}\hat{\bar{\mathbf{B}}}^{H}$. Afterwards, we compute the eigen decomposition of $\hat{\bar{\mathbf{C}}} = \hat{\bar{\mathbf{U}}}\boldsymbol{\Sigma_1} \hat{\bar{\mathbf{U}}}^{H}$using the QR implicit iteration method \cite{GolubMatrix},\cite{xu2023fast}. Subsequently, we take the first-column of the right singular matrix, i.e., 
\begin{align}
    \label{eq:0529016}
    \mathbf{V}^{H} = \boldsymbol{\Sigma_1}^{-\frac{1}{2}}\hat{\bar{\mathbf{U}}}^{H}\hat{\bar{\mathbf{B}}},
\end{align}
followed by performing the element-wise operation by
performing $\mathrm{exp}(j\mathrm{Phase}([\mathbf{V}]_{:,1}))$. We note that the above process avoids the extra computation cost of multiplication of $(\hat{\bar{\mathbf{H}}} \circ \hat{\bar{\mathbf{G}}}^{} )\boldsymbol{\varpi}_{\tau}$ in place of $\hat{\bar{\mathbf{H}}} \circ \hat{\bar{\mathbf{G}}}^{}$. 
The scalability of the above low-complexity method will be analyzed in the following Subsection \ref{sub:scal}.




\subsection{Distance-Inspired Multi-RIS-User/BS Association}
\label{sub:asso}
Thus, the SINR of the direct links and the RIS reflection links 
with given optimal design of phase-shifts of multiple RISs in (\ref{eq:0035}) can be respectively defined as follows:
\begin{align}
\begin{split}
\label{eq:0038}
&\hat{\bar{\gamma}}_D(r_{\text{SD}})= 
\frac{|h_{d,u,i}|^2\cdot  M_r M_t  P \zeta^2\cdot r_{\text{SD}}^{-\alpha}}{\sigma^2+\sum\limits_{i=2}^{U}  
|h_i |^2\cdot \rho_{t,r} \cdot P\cdot \Omega_{\text{SI}}(\xi_{i})},
\end{split}
\end{align}
and
\begin{equation}
\label{eq:01039}
\begin{split}
&\hat{\bar{\gamma}}_I(r_{\text{SR{n}}} \cdot r_{\text{R{n}D}}):= \\&\frac{\sum_{n=a_{1}}^{a_{K}}\left| R_{n}^{\star} \right|^2  \cdot  M_r M_t  P \zeta^2 
\cdot (r_{\text{S} \text{Rn}}  \cdot r_{\text{R{n}D}})^{-\alpha}}{\sigma^2+\sum\limits_{i:\dot{Y}_{i}\in\Phi_R}  
|h_i|^2 \rho_{t,r} P \Omega_{\text{SI}}(\xi_{i}) +
\sum\limits_{j:\dot{Y}_{j}\in\Phi_R \backslash \dot{Y}_{i}}
|h_j |^2 \rho_{t,r}  P \Omega_{\text{SI}}(\xi_{j})},
\end{split}
\end{equation}
where $r_{\textrm{SD}}$, $r_{\text{SR{n}}}$, and $r_{\text{R{n}D}}$ represent the spatial distance from the source to TU, the distance from source to RIS-$n$, and the distance from RIS-$n$ to TU, respectively. \footnote{
Note that, performance analysis of distributed RISs-assisted large networks taking into account the inter-cell interference in multi-cell dense deployments using SG will be considered and followed in our next step research.}   
It is noted that user is interfered not only by the IUs through the selected RIS (i.e., alive RISs) reflective link $\dot{Y}_i \in \Phi_R$, but also by the IUs through the unselected RIS (i.e.,  idle RISs) link $\dot{Y}_{j} \in \Phi_R$.
Besides, $[\hat{\bar{\underline{\boldsymbol{\omega}}}}^{\star}]_{l}$ corresponds to the $l$-th reflecting element response of RISs selected in the set $\mathcal{K}$ using the optimal phase-shifting control criterion in (\ref{eq:0035}).

By following the SINR of the direct links and RIS reflection links derived in (\ref{eq:0038}) and (\ref{eq:01039}), we observe that the physical distance among the users, BS, and distributed RISs impose significant influence on the E2E SINR, i.e., $\hat{\bar{\gamma}}_I(r_{\text{SR{n}}} \cdot r_{\text{R{n}D}}) + \hat{\bar{\gamma}}_D(r_{\text{SD}})$. More concretely, when enlarging the RIS density $\lambda_{R}$, the TU will be likely associated with a set of nearby RISs that provide the strongest directivity gain of each E2E channel link, i.e, making the numerator of $\hat{\bar{\gamma}}_I(r_{\text{SR{n}}} \cdot r_{\text{R{n}D}})$ in (\ref{eq:01039}) as large as possible, thus enabling a maximum E2E SINR at the TU. However, when using all $N$ RISs to assist the dual-hop mmWave communications, i.e., the interference of $\hat{\bar{\gamma}}_I(r_{\text{SR{n}}} \cdot r_{\text{R{n}D}})$ in (\ref{eq:01039}) $\sum\nolimits_{i:\dot{Y}_{i}\in\Phi_R}  
|h_i|^2\cdot \rho_{t,r}  P \Omega_{\text{SI}}(\xi_{i})$ will become enlarged, as the directional beamforming gain $\rho_{t,r}$ will become the main-lobe directivity gain caused by the extremely high RIS density, thus degrading the E2E SINR at the TU. In this case, 
we deduce that the association between the RISs and users is critical to improve the E2E SINR performance. 
With given optimal design of phase shifts of multi-RISs, we come up with the following  distance-inspired dynamic multi-RIS-user association criterion:
\begin{equation}
\label{eq:0036}
\{a_{1}^{\star}, \ldots,a_{K}^{\star}\} = \arg\max_{I\in\{1,\ldots, \binom{N}{K}\}} \hat{\bar{\gamma}}_I(r_{\text{SR{n}}} \cdot r_{\text{R{n}D}}) + \hat{\bar{\gamma}}_D(r_{\text{SD}}).
\end{equation}


\subsection{Directional Beamforming Using MRT}
Once the dynamic multi-RIS-user association and the adaptive phase shifter design are determined, the optimal directional beamforming $\mathbf{w}$ can be achieved by the MRT based on the SINR of the direct links and the RIS reflection links in (\ref{eq:0038}) and (\ref{eq:01039}).   
Given the received SNR from (4), it is well known that the
maximum SNR can be achieved by using a MRT beamformer, i.e.,
\begin{align}
\label{eq:0043}
\mathbf{W}^{\star}=\frac{\hat{\bar{\mathbf{G}}}  \text{bdiag}(\boldsymbol{\Theta}_{a^{\star}_{1}},\ldots, \boldsymbol{\Theta}_{a^{\star}_{K}}) \hat{\bar{\mathbf{H}}}^{T} +\mathbf{H}_{d}}{\left\|\hat{\bar{\mathbf{G}}}  \text{bdiag}(\boldsymbol{\Theta}_{a_{1}},\ldots, \boldsymbol{\Theta}_{a_{K}}) \hat{\bar{\mathbf{H}}}^{T}+\mathbf{H}_{d}\right\|_{\mathrm{F}}^{}}.
\end{align}

\subsection{Scalability of Proposed Phase-Shifting Control Strategies}
\label{sub:scal}

The computational complexity of the proposed phase-shift control strategy is incurred mainly by the SVD operation on $\hat{\bar{\mathbf{H}}} \circ \hat{\bar{\mathbf{G}}}^{}$ in (\ref{eq:0035}).
Specifically, the computational complexity order of the proposed phase-shifting control in Subsection \ref{suc:opti} is \(\mathcal{O}(K^3N_{u}N_{b}L_n^{2})\). In addition, the computational complexity order of the sub-optimal phase-shifting control strategy in Subsection \ref{suc:subo} is \(\mathcal{O}(\tau K^{2}N_{u}N_{b}L_n)\) with \(\tau\) being the rank of the matrix $\hat{\bar{\mathbf{H}}} \circ \hat{\bar{\mathbf{G}}}^{}$. 
However, the computational complexity of the exhaustive RIS-aided scheme proposed in \cite{9205879} involves an \(\mathcal{O} (\prod_{n=1}^{N} b_{n}^{L_{n}})\) complexity cost, with $b_n$ being the number of quantization bits for RIS $n$.

On the other hand, the dynamic RIS-user/BS association criterion mentioned in Subsection \ref{sub:asso} could result in considerable signaling overhead and latency in real-world scenarios. In addition, the update of the RIS-user/BS association criterion depends mainly on the coherence time as well as the update speed of the channel varying.

\section{Performance Analysis}
\label{sec:pa}
In this section, theoretical performance analysis is conducted by leveraging the inherent void probability, contact distribution of the PPP, and the LoS probability \cite{MOLTCHANOV20121146}. At the same time, we also study the PDF of the distance between the TU and its associated RIS to get the location constraints of its associated RIS, and then obtain the location distribution of its associated RIS.

\textbf{Assumption 3:}
(\textit{LoS Probability}) For the LoS probability\cite{docomo20165g}, we use 3GPP model to represent it and thus derive the probability of LoS link as (\ref{eq:09}), which is shown on the top of the next page\cite{docomo20165g}.\footnote{Note that the effect of blockage modeling on signal transmission in direct LoS link has been considered in the LoS probability $\mathrm{Pr}_{\mathrm{LoS}}(\xi)$ derived by Eq. (\ref{eq:09}). How to accurately model the spatial distribution of blockage in distributed multi-RIS-aided large networks, taking into account transient signal blockage effects caused by dynamic obstructions, will be studied in our next step research.} In (\ref{eq:09}), $h_{\mathrm{UT}}$ refers to the actual antenna height at the user terminal,
\begin{figure*}[t]
\begin{align}
\small
\label{eq:09}
\begin{split}
\mathrm{Pr}_{\mathrm{LoS}}(d_{2\mathrm{D}})=\begin{cases}1,&\;d_{2\mathrm{D}}\leq18\mathrm{m},\\\left[\frac{18}{d_{2\mathrm{D}}}+\exp\left(-\frac{d_{2\mathrm{D}}}{63}\right)\left(1-\frac{18}{d_{2\mathrm{D}}}\right)\right]\left(1+\bar{C}(h_{\mathrm{UT}})\frac{5}{4}\left(\frac{d_{2\mathrm{D}}}{100}\right)^{3}\exp\left(-\frac{d_{2\mathrm{D}}}{150}\right)\right),&\;d_{2\mathrm{D}} >18\mathrm{m}, \end{cases}
\end{split}
\end{align}
\hrulefill
\end{figure*}
and $\bar{C}(h_{\mathrm{UT}})$ is equal to
\begin{equation}
\label{eq:10}
\bar{C}(h_{\mathrm{UT}})=\begin{cases}0,&~h_{\mathrm{UT}}\leq13\,\mathrm{m},\\\left(\frac{h_{\mathrm{UT}}-13}{10}\right)^{1.5},&~13\,\mathrm{m}<h_{\mathrm{UT}}\leq23\,\mathrm{m}.\end{cases}
\end{equation}

\begin{lemma}
(\textit{RISs Assist Probability}) Since we assume that there is always a LoS link between the BS and RISs, the probability that the RIS can provide a reflection link for the BS and the user is $\mathrm{Pr}_{\mathrm{LoS}}(r)$.
By following the location-dependent thinning theory \cite{6932503} and with the given RIS PPP $\Phi_{R}$ and the probability $\mathrm{Pr}_{\mathrm{LoS}}(r)$, we can get an inhomogeneous PPP $\Phi_R^L$ with a density of 
\begin{align}
\label{eq:0013}
\lambda_{R}^{L}=\lambda_{R}\cdot \mathrm{Pr}_{\mathrm{LoS}}(r).
\end{align}
\end{lemma}

For the SINR of the direct link, we can calculate it by using the formula (\ref{eq:009}). For the reflective links, we determine the desired RIS through the following association criterion. The user communicates with its serving BS through the link providing the highest E2E SINR. To be concrete, the NLoS link, i.e., the BS-RIS $n$-user link, providing an E2E SINR that higher than that of the rest $N-1$ RIS reflective links, is chosen as the optimal RIS reflective link. In this case, the SINR received at user adopting the optimal RIS reflective link can be expressed as
\begin{equation}
\label{eq:0015}
\gamma^{\star} := \arg\max_{I\in\{1,\ldots,N\}} \gamma_I + \gamma_D.
\end{equation}
It is noted that, by following the above criterion in (\ref{eq:0015}), we can determine the location of the strongest RIS as well. When taking the scenario of distributed RISs into account, the corresponding SINR received at the user
adopting a set of $\mathcal{K}$ selected RIS reflective links yields
\begin{equation}
\label{eq:000021}
\hat{\bar{\gamma}}^{\star} :=\arg\max_{I\in\{1,\ldots, \binom{N}{K}\}} \hat{\bar{\gamma}}_I + \hat{\bar{\gamma}}_D.
\end{equation}

\begin{lemma}
Given that the TU observes at least one LoS RIS, the conditional PDF of its distance to the nearest LoS RIS is\cite{6932503}
\begin{equation}
\label{eq:016}
f_L(x)=2\pi\lambda_Rx\mathrm{Pr}_{\mathrm{LoS}}(x)\mathrm{e}^{-2\pi\lambda_R\int_0^xr\mathrm{Pr}_{\mathrm{LoS}}(r)\mathrm{d}r}/B_L,
\end{equation}
where $x > 0$, and $B_L$ denotes the probability
that a user has at least one LoS RIS and is equivalent to $P_R^s(R_0)$ as described later. $\mathrm{Pr}_{\mathrm{LoS}}(r)$
is the LoS probability function as defined in (\ref{eq:09}).
\end{lemma}

\begin{lemma}
The probability that the TU is associated with a LoS RIS is equal to
\begin{equation}
\label{eq:017}
A_L = B_L\int_0^\infty \mathrm{e}^{-2\pi\lambda_R\int_0^{\psi_{L}(x)} (1-\mathrm{Pr}_{\mathrm{LoS}}(t))t\mathrm{d}t} f_{L}(x) \mathrm{d}x,
\end{equation}
where the term $\psi_{L}(x) = (C_{N}/C_{L})^{1/\alpha_{N}} x^{\alpha_{L}/\alpha_{N}}$. In addition, $\alpha_{L}$ and $\alpha_{N}$ are the LoS and NLoS path-loss exponents, respectively, and $C_{L},C_{N}$ are the intercepts of the LoS and NLoS path-loss formulas, respectively.   
\end{lemma}

Furthermore, conditioning on that the serving BS is LoS (or NLoS), the distance from the TU to its serving BS follows the distribution given in the following lemma.

\begin{lemma}
Assuming that the user is associated with the LoS BS and following the PDF of distance of user and its desired RIS in Lemma 2, the PDF of the distance to its serving BS can be formulated as follows:
\begin{equation}
\label{eq:018}
\hat{f}_{L}(x)=\frac{B_{L}f_{L}(x)}{A_{L}}\mathrm{e}^{-2\pi\lambda_R\int_0^{\psi_{L}(x)}(1-\mathrm{Pr}_{\mathrm{LoS}}(t))t\mathrm{d}t}.
\end{equation}
Thus, we obtain the expected distance from the TU to the associated RIS according to expectation formula $\mathbb{E}[X]=\int_0^\infty\mathrm{x} \hat{f}_{L}(x) \mathrm{dx}$.
\end{lemma}


\begin{lemma}
 Provided that the distance between the user and the BS is known, it is referred to as $\xi$. Then, the probability of existing at least one RIS reflective link within the cell is equal to
\begin{equation}
\begin{split}
\label{eq:019}
&P_R^s(\xi)=
\\&1-\exp\Big(-\int_0^{2\pi}\lambda_R^L\frac{\big(\sqrt{R^2-\xi^2\sin^2\dot{\psi}}-\xi\cos\dot{\psi}\big)^2}{2}\mathrm{d}\dot{\psi}\Big).
\end{split}
\end{equation}
\end{lemma}
\begin{IEEEproof}
In the formula (\ref{eq:019}), we first calculate the number of RISs capable of providing reflective LoS links. Here we let $\dot{r}$ denote the distance between the user and the RIS that assists the user in communication, and $\dot{\psi}$ denote the supplementary angle of the RIS-TU link and the BS-TU link. By resorting to the law of cosine, we replace $\dot{r}$ with the following expression, i.e.,
\begin{align}
\label{eq:020}
\dot{r} =\sqrt{R^{2}-\xi^{2}\sin^{2}\dot{\psi}}-\xi\cos\dot{\psi}.
\end{align}
Hence, the number of RISs that can provide reflective LoS links can be derived as \cite{gan2024coverage} 
\begin{equation}
\label{eq:021}
\begin{aligned}
L^{s}(\xi)& =\int_0^{2\pi}\lambda_R^L\frac{\dot{r}^{2}\mathrm{d}\dot{\psi}}2 \\
&=\int_{0}^{2\pi}\lambda_{R}^{L}\frac{\big(\sqrt{R^{2}-\xi^{2}\sin^{2}\dot{\psi}}-\xi\cos\dot{\psi}\big)^{2}}{2}\mathrm{d}\dot{\psi}.
\end{aligned}
\end{equation}
Then, the corresponding probability can be expressed as
\begin{equation}
\label{eq:022}
\begin{aligned}&P_{R}^{s}(\xi)=1-\exp(-L^{s}(\xi))\\&=1-\exp\Big(-\int_0^{2\pi}\lambda_R^L\frac{\big(\sqrt{R^2-\xi^2\sin^2\dot{\psi}}-\xi\cos\dot{\psi}\big)^2}{2}\mathrm{d}\dot{\psi}\Big),
\end{aligned}
\end{equation}
which completes the proof.
\end{IEEEproof}

Define the coverage probability as
\begin{equation}\label{eq:023}
P_{cov}(T)=\mathrm{Pr}(\gamma>T),
\end{equation}
where $T$ refers to the preset threshold, and $\gamma$ denotes the SINR of the associated link, which could be either a direct link or a reflective link.

\begin{theorem}
(\textit{Ergodic Coverage Probability}) The ergodic
coverage probability of the cell is equal to
\begin{equation}
\label{eq:024}
\mathbb{E}[P_{cov}^s(T)]:=\frac{\int_0^RP_{cov|\xi}^s(T)\lambda_u(\xi)2\pi\xi\mathrm{d}\xi}{\int_0^R\lambda_u(\xi)2\pi\xi\mathrm{d}\xi},
\end{equation}
where $\lambda_u$ is the density of users.
\end{theorem}
\begin{IEEEproof}
See Appendix A.
\end{IEEEproof}

Since the density of users is known, the calculation can be conducted via this integral. We attempt to study the influence of the density of RISs $\lambda_{R}$ on the ergodic coverage probability $\mathbb{E}[P_{cov}^s(T)]$ and find out in our simulation that the density of RISs $\lambda_{R}$ has a direct impact on $\mathbb{E}[P_{cov}^s(T)]$.

\begin{lemma}
Given the distance between the user and its serving BS is $\xi$, then the direct association probability is equal to
\begin{equation}
\label{eq:025}
\mathrm{Pr}_{\mathrm{Ad}}(\xi)=\mathrm{Pr}_{\mathrm{LoS}}(\xi).
\end{equation}
\end{lemma}

\begin{lemma}
Given the distance between the user and its serving BS is $\xi$ and following Lemma 5, the probability that the serving BS and the user is associated through the reflective LoS link is equal to
\begin{equation}
\label{eq:026}
\mathrm{Pr}_{\mathrm{AI}}(\xi)=(1-\mathrm{Pr}_{\mathrm{LoS}}(\xi))P_R^s(\xi).
\end{equation}
\end{lemma}

\begin{IEEEproof}
In Lemma 7, $(1-\mathrm{Pr}_{\mathrm{LoS}}(\xi))$ represents the probability of the direct link between the BS and the user is NLoS. Under this condition, $P_{R}^{s}(\xi)$ is the probability of that there exists a reflective LoS link between the TU and the BS. The proof is completed.
\end{IEEEproof}

We use the coefficient weighting method to get the new SINR $\gamma_{\text{new}}$ and compare it with the threshold $T$. Then, based on the Lemma 7 and the direct link SINR in (\ref{eq:009}) and the RIS reflection link SINR in (\ref{eq:0100}), we have
\begin{equation}
\label{eq:027}
\gamma_{\text{new}}=\mathrm{Pr}_{\mathrm{LoS}}(\xi)\gamma_{D}+(1-\mathrm{Pr}_{\mathrm{LoS}}(\xi))P_R^{s}(\xi)\gamma_{I}.
\end{equation}
Define $P_{cov|\xi}^{s}(\mathrm{T})$ as the  conditional coverage probability of users at a distance $\xi$ from BS. With a given preset SINR threshold $T$, its expression can be expressed as
\begin{align}
\begin{split}
\label{eq:0028}
P_{cov|\xi}^{s}(\mathrm{T}) :=& \mathrm{Pr}_{\mathrm{LoS}}(\xi) \cdot P_{{cov_{d}|\xi}}^{s}(T)+\\
 &\quad\quad (1-\mathrm{Pr}_{\mathrm{LoS}}(\xi))\cdot P_{R}^{s}(\xi)\cdot P_{{cov_{I}|\xi}}^{s}(T),
\end{split}
\end{align}
where $P_{cov_d|\xi}^{s}(T)$ denotes the conditional coverage probability of the direct links and can be expressed as follows
\begin{equation}
\label{eq:0029}
\begin{split}
&P_{cov_d|\xi}^{s}(T)=\mathrm{Pr}(\gamma_D>T)\\
&=\Pr\left(\left|h_{d,u,i}\right|^2>\left(\sigma^2+\text{Interference}\right)\frac T{M_rM_t\cdot P\cdot\Omega_{\mathrm{SD}}(\xi)}\right)\\
& = 1- F_{|h_{d,u,i}|^{2}} \left( \left(\sigma^2+\text{Interference}\right)\frac T{M_rM_t\cdot P\cdot\Omega_{\mathrm{SD}}(\xi)}\right)\\
&\overset{(\text{II})}{\approx} 1-F_{h_{d,u,i}}\left(\left(\frac{\sigma^2T}{M_rM_t\cdot P\cdot \zeta \cdot \xi^{-\alpha}}\right)^{\frac{1}{2}}\right),
\end{split}
\end{equation}
in which the step (II) in (\ref{eq:0029}) is based on the property of the association criterion given in Lemma 1. Specifically, the interference part of the channel owns a lowest directional beamforming gain $\rho_{t,r}$ with a high probability of $ (1-\frac{\psi_t}{2\pi})\cdot (1-\frac{\psi_r}{2\pi})$. Thus, it is believed that the interference is negligible and can be eliminated. Note that a
similar assumption is also applied in the works \cite{xu2024stochastic},\cite{9558795}.
Besides, the CDF expression of the BS-TU channel $|h_{d,u,i}|$, $u\in\{1,\dots,N_{u}\}$, $i\in\{1,\ldots,N_{b}\}$, can be expressed as
\begin{equation}
\label{eq:00290}
F_{h_{d,u,i}}(x_{0})=\int_0^{x_{0}}\frac{t^{m_{h_{d,u,i}}-1}e^{-t/\eta}}{\eta^{m_{h_{d,u,i}}}\Gamma(m_{h_{d,u,i}})}dt,
\end{equation}
where $\eta = \Omega _{h_{d,u,i}}/m_{h_{d,u,i}}$. $\Gamma(m_{h_{d,u,i}}) =\int_0^\infty t^{m_{h_{d,u,i}}-1}e^{-t} dt $ refers to the Gamma function. Likewise, the PDF expression of the BS-TU channel $|h_{d,u,i}|$ can be expressed as
\begin{equation}
f_{h_{d,u,i}}(x_{0})=
\frac{x_{0}^{m_{h_{d,u,i}}-1}e^{-x_{0}/\eta}}{\eta^{m_{h_{d,u,i}}} \Gamma(m_{h_{d,u,i}})}.
\end{equation}

Likewise, the conditional coverage probability of the reflective LoS link is equal to
\begin{equation}
\label{eq:00310}
P_{cov_I|\xi}^{s}(T) = \mathrm{Pr}(\gamma_I>T).
\end{equation}
To derive the above conditional coverage probability $P_{cov_I|\xi}^{s}(T)$ given in (\ref{eq:00310}),
we first let $R_{n} =\sum_{l=1}^{L_n}h_{n,l}\left[\underline{\omega}^\star\right]_l g_{n,l} = \sum_{l=1}^{L_n} U_{n,l}$.  
With a given reset SINR threshold $T$, we can calculate the coverage of the reflective link in (\ref{eq:032}), as shown on the top of next page, where $\dot{r}_{n}$ denotes the distance between the user and the $n$-th RIS, and $\dot{s}_{n}$ represents the distance between the BS and the $n$-th RIS. 
\begin{figure*}
\begin{align}
\small
\label{eq:032}
\begin{split}
\mathrm{P}_{{\mathrm{cov}_{{\mathrm{I}}}|\xi}}^{{\mathrm{S}}}(\mathrm{T}) &= \mathrm{Pr}\left(\begin{array}{c}\left|\sum_{l=1}^{{L_{n}}}h_{n,l}\cdot\left[\underline{\omega}^{\star}\right]_{l}\cdot g_{n,l}\right|^{2}>\left(\sigma^{2}+\mathrm{Interference}\right)\cdot\end{array} \frac{T}{M_{r}M_{t}\cdot P \cdot \Omega_{{\mathrm{SRn}}}(\xi)\cdot\Omega_{{\mathrm{RnD}}}(\xi)}\right)\\
& = \mathrm{Pr}\left(|R_{n}|^2 > \frac{\left(\sigma^{2}+\mathrm{Interference}\right) \cdot T}{M_{r}M_{t}\cdot P \cdot \Omega_{{\mathrm{SRn}}}(\xi)\cdot\Omega_{{\mathrm{RnD}}}(\xi)}\right)\\
& = \mathrm{Pr} \left(\begin{array}{c}
 \dot{r}_{n}^{\alpha}\cdot
 \dot{s}_{n}^{\alpha}
<\end{array} \frac{M_{r}M_{t}\cdot P\cdot \zeta^2 \cdot |R_n|^2 }{\left(\sigma^{2}+\mathrm{Interference}\right)\cdot T}\right),
\end{split}
\end{align}
\end{figure*}
Since both $h_{n,l}$ and $g_{n,l}$ are RVs, thus $R_n$, $n=1,\ldots,N$, are i.n.i.d. RVs. The approximate distribution of $R_n$ can be obtained as $R_n\stackrel{\text{approx.}}{\sim}\mathrm{Gamma}(L_n\alpha_{U_n},\beta_{U_n})$, where $\alpha_{U_n}$ and $\beta_{U_n}$ can be expressed as \cite{9558795}
\begin{align}
\alpha_{U_n}=\frac{\left(\mathbb{E}[U_{n,l}]\right)^2}{\mathrm{Var}[U_{n,l}]}=\frac{[\mu_{U_{n,l}}(1)]^2}{\mu_{U_{n,l}}(2)-[\mu_{U_{n,l}}(1)]^2},
\end{align}
\begin{align}
\beta_{U_n}=\frac{\mathbb{E}[U_{n,l}]}{\mathrm{Var}[U_{n,l}]}=\frac{\mu_{U_{n,l}}(1)}{\mu_{U_{n,l}}(2)-[\mu_{U_{n,l}}(1)]^2},
\end{align} 
where $\mu_{U_{n,l}}(1)$ and $\mu_{U_{n,l}}(2)$ correspond to the first and second moments of the RV $U_{n,l}$, respectively.
We use the $M$-staircase approximation to approach the averaged RV $U_{n,l}$, i.e., $\mathbb{E}[U_{n,l}]$. Then, the $m$-th moment of $U_{n,l}$, $m \in \{1,\ldots,M\}$, can be obtained as
\begin{align}
\mu_{U_{n,l}}(m)=\bar{\eta}_{nl}^{-m}\frac{\Gamma(m_{h_{n,l}}+\frac{m}{2})\Gamma(m_{g_{n,l}}+\frac{m}{2})}{\Gamma(m_{h_{n,l}})\Gamma(m_{g_{n,l}})},
\end{align} where $\bar{\eta}_{nl}=\sqrt{\frac{1}{|\left[\underline{\omega}^\star\right]_l |^{2}}\frac{m_{{h}_{n,l}}}{\Omega_{h_{n,l}}}\frac{m_{g_{n,l}}}{\Omega_{g_{n,l}}}}$. Consequently, the approximate CDF and PDF of $R_{n}$ can be derived by
\begin{align}F_{R_n}(z) \approx\frac{\gamma(L_n\alpha_{U_n},\beta_{U_n}z)}{\Gamma\left(L_n\alpha_{U_n}\right)},\end{align}
\begin{align}\label{eq:00037}
f_{R_n}(z) \approx \frac{\beta_{U_n}^{
L_n\alpha_{U_n}}}{\Gamma(L_n\alpha_{U_n})}z^{L_n\alpha_{U_n}-1}e^{\beta_{U_n} z},z\geq0.
\end{align}
\begin{figure*}
\begin{align}
\small
\label{eq:000331}
\begin{split}
\mathrm{P}_{{\mathrm{cov}_{{\mathrm{I}}}|\xi}}^{{\mathrm{S}}}(\mathrm{T})
& = \mathrm{Pr} \left(
 \mathop{\min}_{1 \leq n\leq N}\left(\dot{r}_{n}\cdot
\dot{s}_{n}\right)
< \left(\frac{M_{r}M_{t}\cdot P \cdot\zeta^2 \cdot |R_n|^2}{\left(\sigma^{2}+\mathrm{Interference}\right) \cdot T}\right)^{\frac{1}{\alpha}}\right)\\
& = \int_{0}^{\infty} F_{\eta_{0}|\xi}\left(\left(\frac{M_{r}M_{t} \cdot P \cdot\zeta^2 \cdot |R_{n}|^2}{\left(\sigma^{2}+\mathrm{Interference}\right)\cdot T}\right)^{\frac{1}{\alpha}}\right)f_{|R_{n}|^2}(x_{})\mathrm{d}x_{}\\
& \approx \int_{0}^{\infty} F_{\eta_{0}|\xi}\left(\left(\frac{M_{r}M_{t} \cdot P \cdot\zeta^2 \cdot  |R_{n}|^2}{\sigma^{2}\cdot T}\right)^{\frac{1}{\alpha}}\right)f_{|R_{n}|^2}(x_{})\mathrm{d}x_{}\\
& \approx \int_{0}^{\infty} F_{\eta_{0}|\xi}\left(\left(\frac{M_{r}M_{t} \cdot P \cdot\zeta^2 \cdot  |R_{n}|^2}{\sigma^{2}\cdot T}\right)^{\frac{1}{\alpha}}\right)f_{R_{n}}(x^{\frac{1}{2}})\mathrm{d}x_{},
\end{split}    
\end{align}    
\hrulefill
\end{figure*}

Define the minimum path length product from the user
to the reflective NLoS RIS $n$ and from the BS to the $n$-th RIS as $\eta_0 = \arg\min_{n\in\{1,\ldots,N\}}\dot{r}_{n}\cdot \dot{s}_{n}$.
Based on the BS-RIS $n$-TU association criterion given in (\ref{eq:0015}) and using the $\mathrm{Gamma}(L_n\alpha_{U_n},\beta_{U_n})$ to approximate the reflective links $R_{n}$, the conditional coverage probability of the reflective LoS
link can be further derived as (\ref{eq:000331}), shown on the top of this page.
\begin{figure*}
\begin{align}
\small
\label{eq:00033}
\begin{split}
F_{\eta_n|\xi}(x)=\begin{cases} 0,&\quad x\le \dot{r}_{n},\\ 1-\exp(-\int_0^R\int_0^{\mathrm{arccos}\big(\frac{\dot{s}_{n}^{4}+\dot{s}_{n}^{2}\xi^{2}-x^{2}}{2 \dot{s}_{n}^{3} \xi}\big)}\lambda_R \mathrm{Pr}_{\mathrm{LoS}}(\sqrt{\dot{s}_{n}^2+\xi^2-2\dot{s}_{n}\xi\cos\dot{\theta}})\dot{s}_{n}\mathrm{d}\dot{\theta}\mathrm{d}\dot{s}_{n}),&\quad\text{otherwise},\end{cases}
\end{split}
\end{align}
\hrulefill
\end{figure*}
Given the BS-TU distance $\xi$, the CDF expression of the distance of BS-RIS-$n$ link and the distance of the RIS-$n$-TU link \cite{gan2024coverage} can be derived as (\ref{eq:00033}),\footnote{Herein, the notation of $\mathrm{arccos}(\cdot)$ function is mainly used to calculate the angle from the given cosine trigonometric ratio.} as shown on the top of next page. In (\ref{eq:00033}), $\dot{\theta}$ indicates the angle between the BS-TU and the BS-RIS-$n$ links. It is noted that the conditional CDF $F_{\eta_n|\xi}(x)$ is derived in (\ref{eq:00033}) by following the intrinsic properties of the PPP, i.e., the contact distribution and the void probability\cite{MOLTCHANOV20121146}, as the location of RISs is distributed by following a  homogeneous
PPP $\Phi_{R}$.

Given the conditional coverage probability $P_{cov|\xi}^s(T)$, the achievable rate of this single-cell can be mathematically expressed as $\Gamma = \mathrm{BW}\cdot \log_{2} \left(1+ \min(\gamma_{\text{new}}, T)\right) $, where $\mathrm{BW}$ denotes the bandwidth allocated to the TU.
Next, we give the expression for the sum rate in the region, and the ergodic rate of a TU at a distance $\xi$ from BS is 
\begin{equation}\begin{aligned}
\Psi(\xi)& =\mathbb{E}\left[\mathrm{BW} \cdot \log_{2}(1+\min(\gamma_{\text{new}}, T))\right]
\\&=\int_{0}^{\infty}P_{cov|\xi}^{s}(2^{\frac{t}{\mathrm{BW}}}-1)\mathrm{d}t.
\end{aligned}\end{equation}
Since the location of each user is known, the sum rate of the region can be expressed as
\begin{equation}
\Psi =\sum_{i=1}^U\int_0^\infty P_{cov|\xi_i}^s(2^{\frac{t}{\mathrm{BW}}}-1)dt,\end{equation}
where $\xi_i$ is the distance from each user to the BS.

\begin{table}[t!]
\linespread{1.3} 
\captionsetup{font={scriptsize,sc}, format=plain, singlelinecheck=off}
\captionsetup{justification=centering}
\tocaption{SYSTEM PARAMETER SETUPS}
  \centering
  \scriptsize
  \begin{tabular}{|l |c|}
    \hline
    \textbf{System Parameters} & \textbf{Values}  \\
     \hline
      \hline
       Carrier Frequency $f_{c}$&  28 GHz   \\
    \hline
     Bandwidth $\mathrm{BW}$& 200 MHz \\
     \hline
     Cellular Radius $R$ & [100m, 250m] \\
     \hline
     Number of Antennas at the BS $N_{b}$ & 64 \\
     \hline
     Number of Users $U$ & \{10, 30, 40, 50, 60\}\\
     \hline
     Number of Antennas at the User $N_{u}$ & [1, 4] \\
     \hline
     Density of RISs, i.e., $\lambda_{R}$ & $[1.5 e{-4}/ \mathrm{m}^{2}, 5.5 e{-3}/ \mathrm{m}^{2}] $ \\
     \hline
     Location Distribution of RISs &  Homogeneous PPP, PCP\cite{10064007,6524460}\\
    \hline
     Antenna Gain $G_{\text{S}}$, $G_{\text{Rn}}$, $G_{\text{D}}$ & 10 dBi, 10 dBi, 5 dBi \\
     \hline
     Small-Scale Fading Channel Model & Nakagami-$m$ Fading\\
     \hline
     Nakagami Scale and Spread Parameters $m$ & $\left(2.5, 1.5\right)$ \\
     \hline
     Transmit Power & 8 dBm \\
     \hline
     Noise Figure & 10 dBm\\
     \hline
     Reference Distance $d_{0}$ & 1 m  \\
      \hline
      Heights of BS and Users & 25 m \&1.5 m   \\
       \hline
      Height of Environment & 1 m  \\
      \hline
  \end{tabular}
\label{Tb:Setting}
\end{table}

%
\section{Simulation Results}
\label{sec:simulations}
\subsection{Simulation Specification}

\begin{figure}[tp]
\setlength{\abovecaptionskip}{-0.1cm}
\setlength{\belowcaptionskip}{-0.4cm}
    \begin{center}
\includegraphics[width=0.47\textwidth]{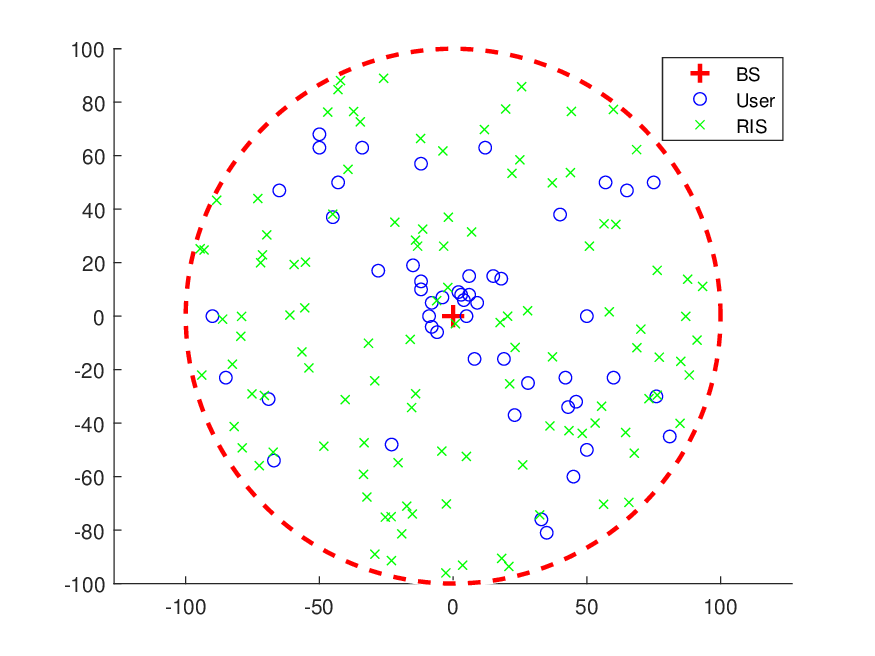}
    \end{center}
    \caption{The spatial location of the BS, RISs and users in a single-cell cellular region with the radius $R = 100 \,m$.} \label{fig:03}
\end{figure}

In this section, we consider the simulation results in two cases: one is to analyze the received signal at the TU and the comprehensive system performance by using the associated RISs determined in \textbf{Lemma 1}; the other is to analyze the impact of distributed RISs in all samples on the SINR coverage and achievable rate performance by using Monte Carlo simulations and calculate the average performance. Note that the spatial location of the BS, RISs and users of one realization in our simulation in a single-cell cellular region is provided in Fig. \ref{fig:03}. Note also that TABLE \ref{Tb:Setting} lists the simulation parameters. These parameters basically follow the 3GPP NR \cite{8458146}. For better evaluating the performance of our proposed scheme, we consider the following benchmark methods:
 \begin{itemize}
 \item \textbf{Exhaustive RIS-aided (ERA) scheme}: all $N$ RISs assist the transmission between the BS and the TU, and all the RISs are controlled to reflect replicas of $x$ to TU over the same time-frequency channel;
 \item \textbf{Quantized phase-shifting scheme}: the number of discrete phase shift is limited and constrained by the phase shift resolution $Q_{n} = 2^{b_{n}}$. The value of the $l$-th phase shift of RIS $n$ is chosen exhaustively from the set $\{0,\frac{2\pi}{Q_{n}},\ldots, \frac{2\pi(Q_{n}-1)}{Q_{n}}\}$ until arriving at a minimum phase error between the unquantized and quantized phase shifts;\footnote{Hereinafter, the quantization bit $b_n$ is set to be 4 for RIS $n$, i.e., 4-bit quantized shifts are considered in the following simulations.}
 \item \textbf{Random phase-shifting scheme}: all $N$ RISs assist the transmission between the BS and the TU, and the discrete phase shift of the $l$-th reflecting element of the $n$-th RIS is set randomly.
 \end{itemize}

\subsection{The Analytical and Simulation Result Analysis}
\begin{figure}[tp]
\setlength{\abovecaptionskip}{-0.1cm}
\setlength{\belowcaptionskip}{-0.4cm}
    \begin{center}
\includegraphics[width=0.47\textwidth]{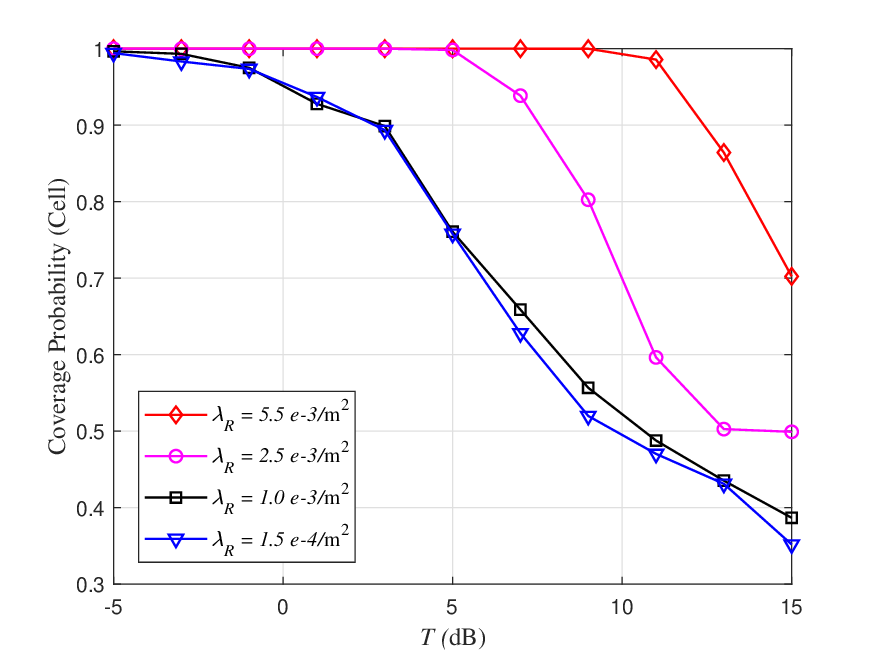}
    \end{center}
    \caption{The SINR coverage probability performance impacted by the SINR threshold for different RIS densities $\lambda_R$ with $N_{b} = 64, U = 30, N_{u} = 4$.} \label{fig:04}
\end{figure}

\begin{figure}[tp]
\setlength{\abovecaptionskip}{-0.1cm}
\setlength{\belowcaptionskip}{-0.4cm}
    \begin{center}
\includegraphics[width=0.47\textwidth]{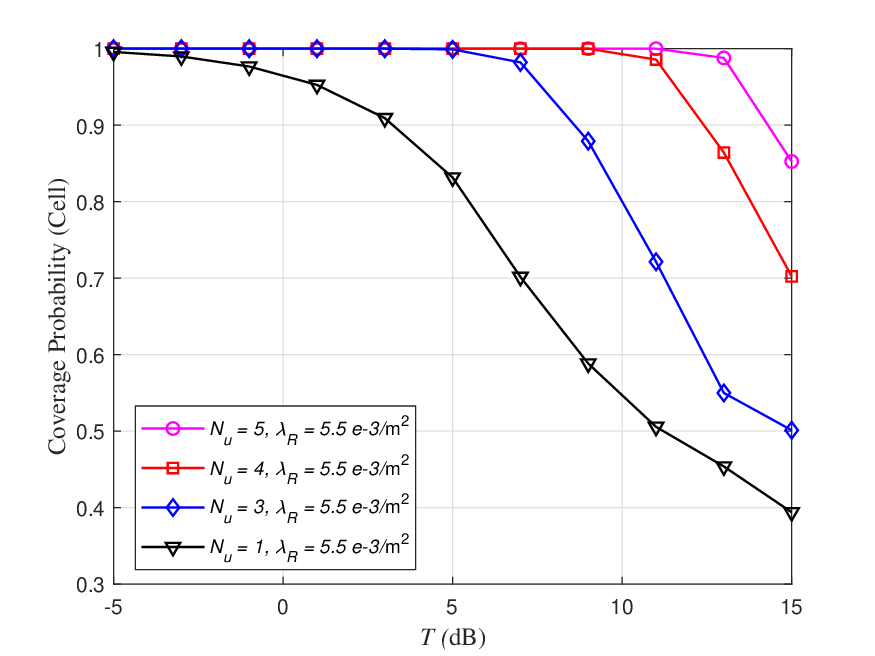}
    \end{center}
    \caption{The SINR coverage probability performance impacted by the SINR threshold for different numbers of UE antennas with $N_{b} = 64$, $U = 30$, and the RIS density $\lambda_{R} = 5.5e{-3} /\mathrm{m}^{2}$.} \label{fig:05}
\end{figure}

Fig. \ref{fig:04} depicts the coverage probability versus the SINR threshold under the varying density of the RISs in /$\mathrm{m}^{2}$ with $N_{b} = 64$ and $N_{u} =4$. It can be seen from Fig. \ref{fig:04} that the SINR threshold can greatly affect the coverage probability, particularly with the scenario of $\lambda_{R} = 1.5 e{-4}/\mathrm{m}^{2}$ (the concrete number of deployed RISs is 5, i.e., $N = 5$). This figure also indicates that with given $T$, increasing the density of RISs leads to an enlarged SINR coverage probability. On the other hand, Fig. \ref{fig:05} shows the coverage probability versus SINR threshold under different antenna configuration at TU with $\lambda_{R} = 5.5 e{-3}/\mathrm{m}^{2}$ ($N=70$). This figure indicates that increasing the number of antennas at the TU increases the coverage probability significantly. The reason is that increasing $N_{u}$ yields a large directional gain $\rho_{t,r}$ as given in (2), thus greatly enlarging the link SINR values in (3) and (4).

\begin{figure}[t!]
\setlength{\abovecaptionskip}{-0.1cm}
\setlength{\belowcaptionskip}{-0.3cm}
    \begin{center}
\includegraphics[width=0.47\textwidth]{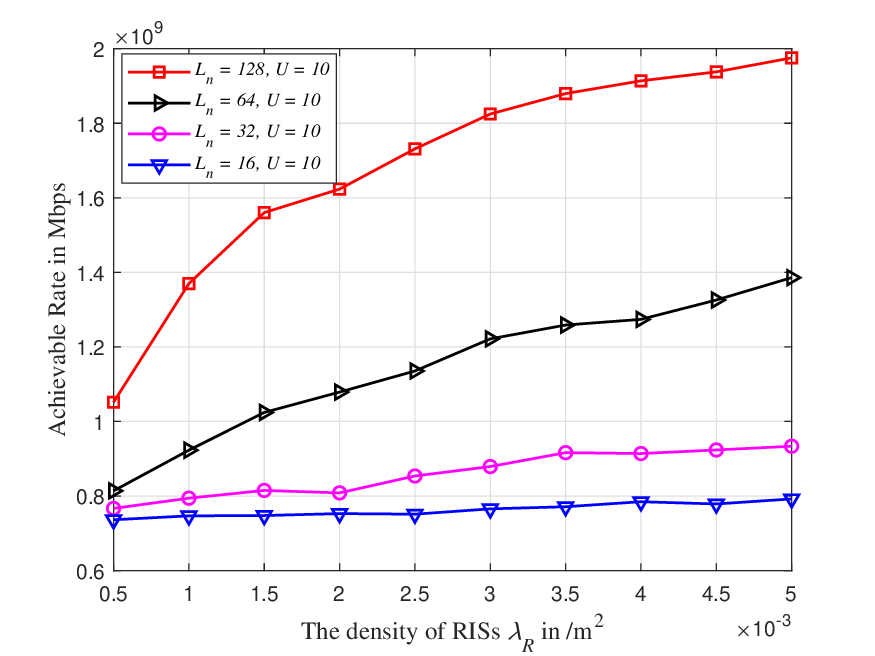}
    \end{center}
    \caption{The achievable rate in Mbps impacted by the RIS density $\lambda_R$ in /$\rm m^{2}$ for different numbers of RIS elements with $N_{b} = 64$, $U = 10$, and the SINR threshold $T = -3 $ dB.} \label{fig:06}
\end{figure}

\begin{figure}[tp]
\setlength{\abovecaptionskip}{-0.1cm}
\setlength{\belowcaptionskip}{-0.4cm}
    \begin{center}
\includegraphics[width=0.47\textwidth]{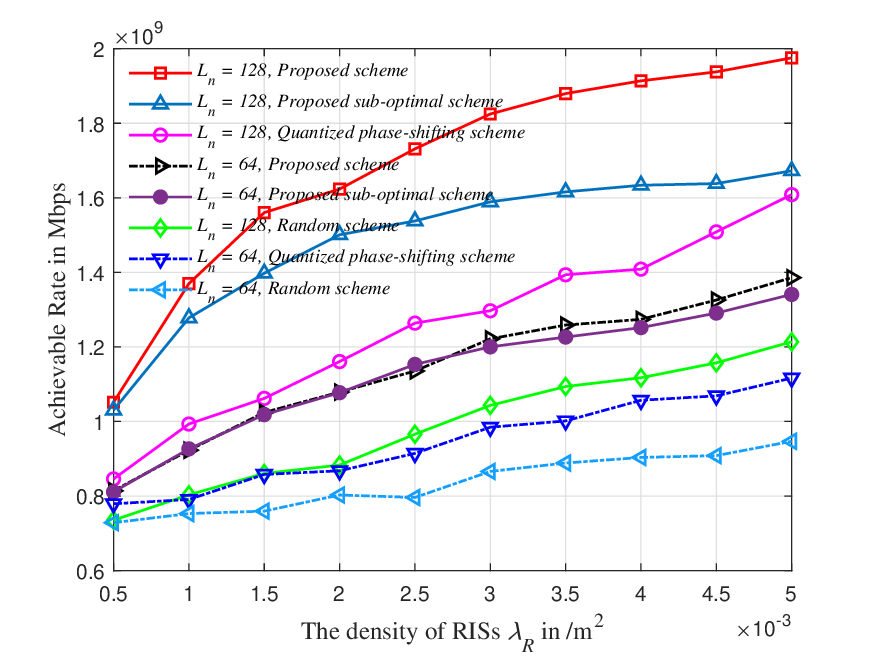}
    \end{center}
    \caption{The achievable rate in Mbps impacted by the RIS density $\lambda_R$ in /$\rm m^{2}$ employing distinct phase-shifting control schemes for different numbers of RIS elements with $N_{b} = 64$, $U = 10$, and the SINR threshold $T = -3 $ dB.} \label{fig:07}
\end{figure}

\begin{figure}[t!]
\setlength{\abovecaptionskip}{-0.1cm}
\setlength{\belowcaptionskip}{-0.3cm}
    \begin{center}
\includegraphics[width=0.47\textwidth]{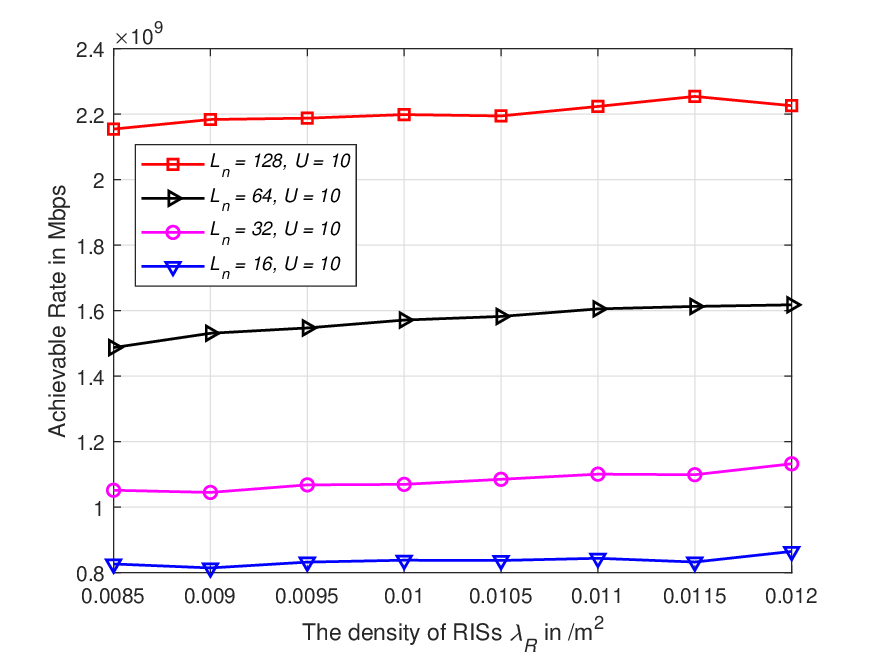}
    \end{center}
    \caption{The achievable rate performance in Mbps impacted by the RIS density $\lambda_R$ in /$\rm m^2$
for different numbers of RIS elements with  $N_b$ = 64, and the SINR threshold $T$ = -3 dB.} \label{fig:0013}
\end{figure}

Fig. \ref{fig:06} portrays the achievable rate in Mbps versus the density of RISs in /$\mathrm{m}^{2}$ for different numbers of RIS passive reflecting elements with $U = 10$ and the SINR threshold $T = -3$ dB. It is easy to see from Fig. \ref{fig:06} that increasing $\lambda_{R}$ yields an increased achievable rate. In addition, with a given achievable rate, configuring larger-sized RISs requires a smaller number of RISs to assist the transmission of data signal from the BS to the TU. Fig. \ref{fig:07} plots the achievable rate in Mbps versus the density of RISs in /$\mathrm{m}^{2}$ employing different phase-shifting control strategies. Basically, this figure illustrates that the proposed optimal phase-shifting control approach based on maximizing the cascaded E2E channel gain enables higher achievable rate than the quantized phase-shifting and random methods with a given RIS density. This is because our approach can increase the degree of freedom of the E2E channel link by optimizing the RIS reflecting elements, thus enhancing the achievable rate performance. In addition, this figure reveals that the proposed sub-optimal phase-shifting control approach can achieve similar rate performance compared to that of the optimal phase-shifting control
method when $L_n = $ 64. However, the performance gap between the above two proposed approaches enlarges when increasing the value of $L_n$.
Fig.~\ref{fig:0013} displays the achievable rate performance impacted by the density of RISs for different numbers of RIS elements with $N_b$ = 64, and the SINR threshold $T$ = -3 dB.
We observe from Figs. \ref{fig:06} and \ref{fig:0013} that: i) the achievable rate performance saturates 
when the RIS density increases from
$8.5e{-3} /\mathrm{m}^{2}$ to $1.2e{-2} /\mathrm{m}^{2}$ due to diminishing link gains brought by additional RISs. ii) With given user density, i.e., $U$, increasing the number of RIS elements $L_n$ leads to significant rate performance improvement.

\begin{figure}[tp]
\setlength{\abovecaptionskip}{-0.1cm}
\setlength{\belowcaptionskip}{-0.3cm}
    \begin{center}
\includegraphics[width=0.47\textwidth]{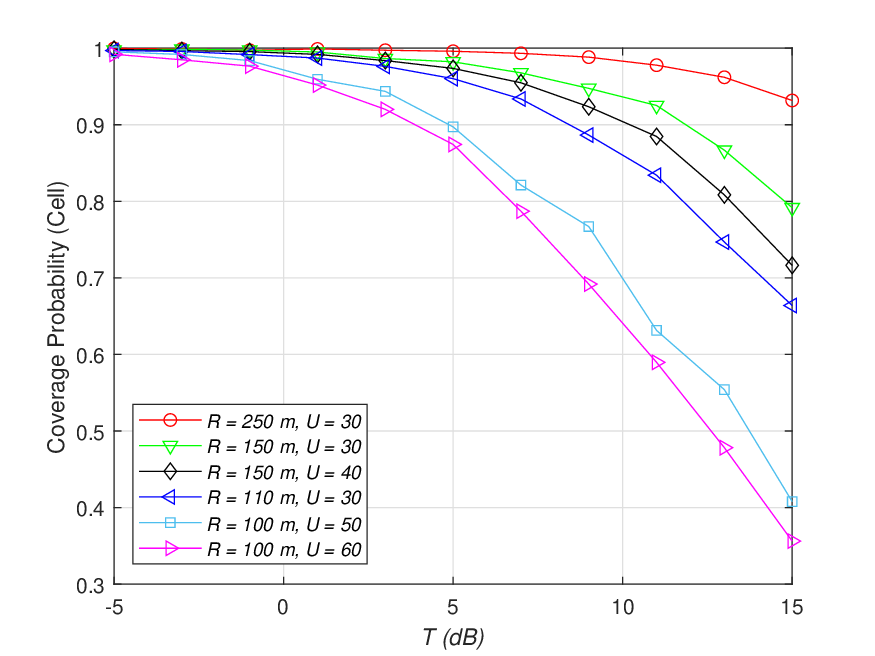}
    \end{center}
    \caption{The SINR coverage probability performance impacted by the SINR threshold under various cellular radius $R$ and user density with $N_b=64$, $N_u=1$, and the RIS density $\lambda_R = 5.5e{-3}$/$\rm m^2$.} \label{fig:0011}
\end{figure}

\begin{figure}[tp]
\setlength{\abovecaptionskip}{-0.1cm}
\setlength{\belowcaptionskip}{-0.3cm}
    \begin{center}
\includegraphics[width=0.47\textwidth]{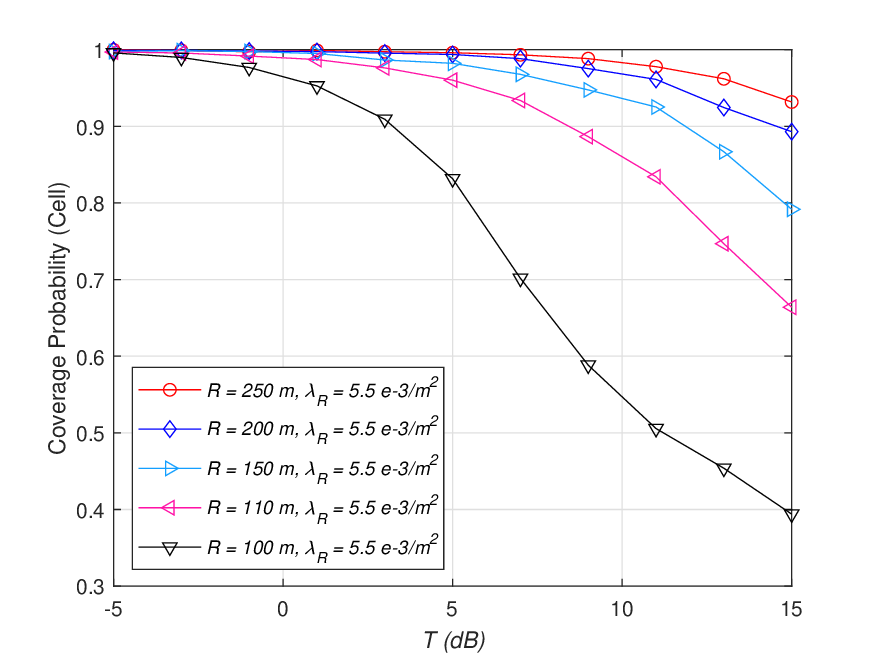}
    \end{center}
    \caption{The SINR coverage probability performance impacted by the SINR threshold under the various cellular radius $R$ with $N_b=64$, $U=30$, and $N_u=1$.} \label{fig:0010}
\end{figure}

In Figs. \ref{fig:0011} and \ref{fig:0010}, we evaluate the coverage probability performance impacted by the SINR threshold under various user density and cellular radius $R$, respectively. Specifically, it is easy to see from Fig. \ref{fig:0011} that with given user density, increasing the value of $R$ yields an improved coverage performance. However, given the value of $R$, enlarging the user density results in degraded coverage probability performance, due to the excessive intra-cell interference originated from the NLoS RISs. On the other hand, it is easy to see from Fig. \ref{fig:0010} that with given SINR threshold $T$, increasing the value of $R$ leads to the coverage performance improvement. The major reason lies in that increasing the value of $R$ will increase the number of RISs deployed in this region.

\begin{figure}[tp]
\setlength{\abovecaptionskip}{-0.1cm}
\setlength{\belowcaptionskip}{-0.3cm}
    \begin{center}
\includegraphics[width=0.47\textwidth]{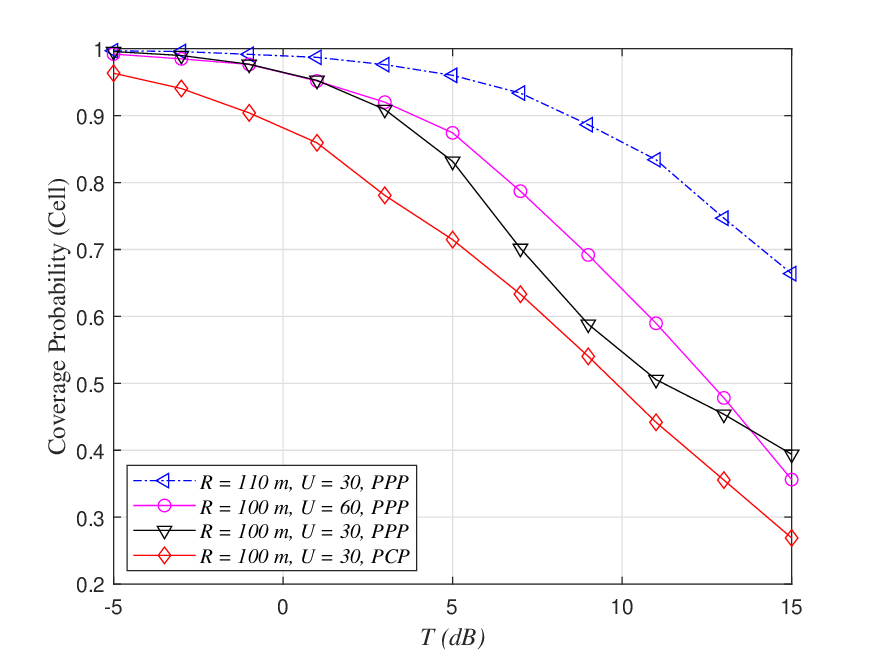}
    \end{center}
    \caption{
The SINR coverage probability performance impacted by the SINR threshold $T$ under various cellular radius $R$, user density and RIS deployment model with $N_b$ = 64, $N_u $ = 1, and the RIS density $\lambda_R = 5.5e{-3}$/$\rm m^2$. Herein, the average number of RISs per cluster is 3, and the standard variance of PCP distribution is set to be 0.0625.} \label{fig:0012}
\end{figure}

Fig. \ref{fig:0012} shows the coverage performance impacted by the SINR threshold $T$ under various cellular radius $R$, user density, and RIS deployment model with $N_b$ = 64, $N_u $ = 1, and the RIS density $\lambda_R = 5.5e{-3}$/$\rm m^2$. In this figure, we adopt the homogeneous PPP and PCP to model locations of RISs, respectively. This figure reveals that with given RIS density $\lambda_R = 5.5e{-3}$/$\rm m^2$, the coverage probabilities of the proposed multi-RIS-assisted dual-hop mmWave networks using PCP and PPP are close to each other. However, when increasing the user density, the proposed multi-RIS-assisted dual-hop mmWave networks using the PPP achieves higher coverage probability.

\begin{figure}[t]
\setlength{\abovecaptionskip}{-0.1cm}
\setlength{\belowcaptionskip}{-0.5cm}
    \begin{center}
\includegraphics[width=0.47\textwidth]{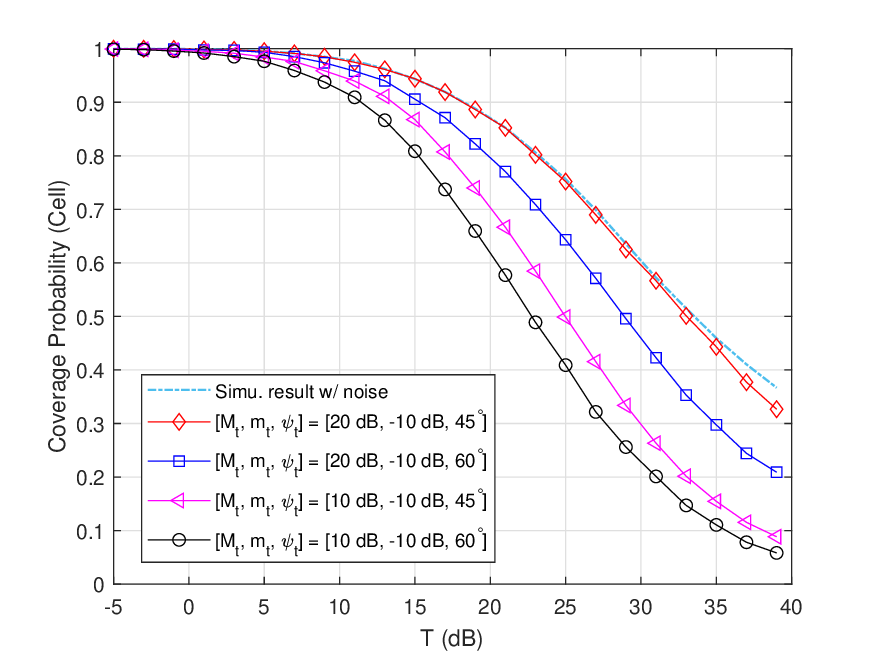}
    \end{center}
    \caption{The simulation  results of the SINR coverage probability impacted by the SINR threshold with different transmit antenna geometry. The receiver beam pattern is configured as [10 dB, -10 dB, $90^{\circ}$], and $N_{b} = 64, U = 10, N_{u} = 1$, and $\lambda_{R} = 1.5e{-3} /\mathrm{m}^{2}$.} \label{fig:08}
\end{figure}

\begin{figure}[t]
\setlength{\abovecaptionskip}{-0.1cm}
\setlength{\belowcaptionskip}{-0.5cm}
    \begin{center}
\includegraphics[width=0.47\textwidth]{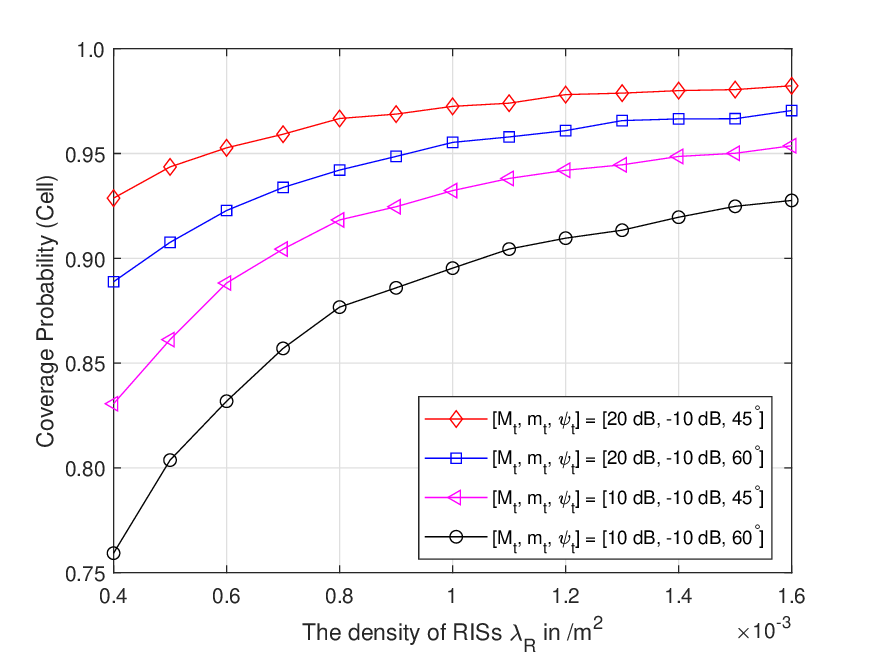}
    \end{center}
    \caption{The simulation results of the SINR coverage probability  impacted by the density of RISs $\lambda_R$ in /$m^{2}$ with different antenna geometry. The beam pattern at the destination side is configured as [10 dB, -10 dB, $90^{\circ}$], and $N_{b} = 64, U = 10, N_{u} = 1$, and the SINR threshold $T = 10$ dB.} \label{fig:09}
\end{figure}

Fig. \ref{fig:08} plots the SINR coverage probability impacted by the SINR threshold under the scenarios of distinct antenna geometry configurations. Herein, we use $[M_t,m_t,\psi_t]$ to 
denote the sectored antenna
pattern at the source side, where 
$M_t$ denotes the main-lobe directivity gain, $m_t$ is the side-lobe gain, and $\psi_t$ means the beamwidth of the main lobe. In addition, in the sectored
antenna model, the array gains are assumed to be constant
$M_t$ for all angles in the main lobe, and another constant $m_t$
in the side lobe in the sectored model.
We observe from this figure that with given $m_t$, $\psi_t$, and the SINR threshold $T$ = 25 dB, increasing the main lobe directivity gain $M_t$ from 10 dB to 20 dB can enable the SINR coverage probability $P_{cov|\xi}^{s}(T)$ performance improvement by
50.7$\%$. In particular, with given $T$ = 35 dB, the SINR coverage probability performance gain is up to 187$\%$. The relevant reason is that the proposed multi-RIS-user association can greatly increase the SINRs at TU by selecting the strongest RIS by (\ref{eq:0015}), thus yielding higher $P_{cov|\xi}^{s}(T)$. In addition, we found out that the beamwidth of the main lobe imposes an influence on $P_{cov|\xi}^{s}(T)$ as well. On the other hand, Fig.  \ref{fig:09} shows the simulation results of SINR coverage probability employing our proposed approach impacted by the density of RISs $\lambda_R$ with different antenna geometry. It is easy to see from this figure that, when the deployment of the RISs is very sparse, and increasing $M_t$ from 10 dB to 20 dB, our approach enables the coverage probability performance gain of 12.7$\%$. In addition, when increasing the density of RISs from $5 e{-4}/\mathrm{m}^{2}$ ($N=17$) to $5 e{-3}/\mathrm{m}^{2}$ ($N=162$), our approach improves the coverage probability by 22.2$\%$. However, we observe from Fig. \ref{fig:09} that when deploying more RISs per cell, the coverage rate gain will become negligible.

\begin{figure}[t]
\setlength{\abovecaptionskip}{-0.1cm}
\setlength{\belowcaptionskip}{-0.3cm}
    \begin{center}
\includegraphics[width=0.47\textwidth]{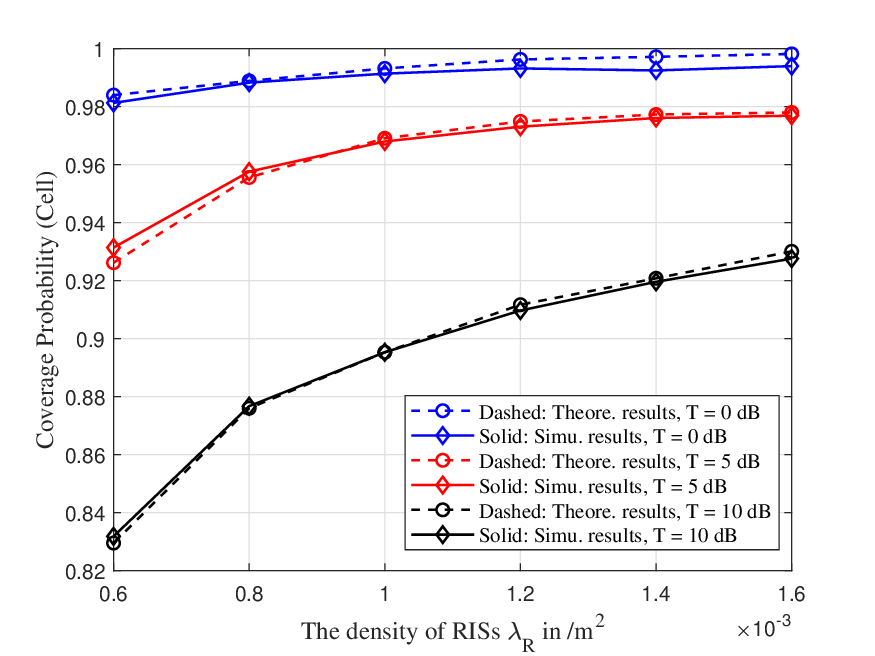}
    \end{center}
    \caption{The theoretical and simulation results of the SINR coverage probability impacted by the density of RISs $\lambda_R$ in /$m^{2}$ with distinct SINR thresholds $T$. 
    The beam pattern at the source side is configured as [10 dB, -10 dB, $60^{\circ}$], whereas the beam pattern at the destination side is configured as [10 dB, -10 dB, $90^{\circ}$], and $N_{b} = 64, U = 10, N_{u} = 1$.} \label{fig:10}
\end{figure}

In Fig. \ref{fig:10}, we evaluate the analytical and simulation results of the SINR coverage
probability employing our approach in multi-RIS-assisted mmWave systems under the scenarios of different RIS density configurations. It is easy to see from this figure that the analytical results in \textit{Theorem 1} are well-aligned to the Monte Carlo simulation results. This is because we use the $M$-staircase approximation to precisely approximate the mean of RVs $U_{n.l} = h_{n,l}\left[\underline{\omega}^\star\right]_l g_{n,l}$, followed by using $\mathrm{Gamma}(L_n\alpha_{U_n},\beta_{U_n})$ to approximate the actual distribution of RIS
$n$-TU channel $R_{n} = \sum_{l=1}^{L_n} U_{n,l}$.    
In addition, we deduce that when using more staircase terms (i.e., large $M$) in the approximation, the approximation will become more accurate. Besides, we observe from this figure that when increasing the SINR thresholds from 0 dB to 10 dB, the coverage probability will be decreased by 18.1$\%$, given that the density of RISs is $6 e{-4}/\mathrm{m}^{2}$ ($N=13$).  

\section{Concluding Remarks}
\label{sec:conclusion}
This paper investigated the effective gain of the E2E SINR coverage probability and achievable rate of distributed multi-RIS-aided mmWave communication system by using stochastic geometry method. Specifically, we
first developed a system model and then analyzed the E2E SINR coverage probability and the achievable rate for the distributed multi-RIS-assisted mmWave communications. Afterwards, to improve the performance in terms of the SINR coverage probability and achievable rate, we optimized the dynamic association criterion and RIS phase-shifting control. 
Furthermore, we optimized the multi-RIS-user association
based on the physical distances between the RISs and the destination, as well as the
phase-shifting control of the RISs. Simulation results indicated that the deployment of distributed RISs can significantly improve the E2E coverage probability and achievable rate of the system compared to the selected benchmarks.
\begin{appendices}
\section{Proof of Theorem 1.}
\label{sec:Proof1}
The conditional SINR coverage probability in the theoretical case consists of the conditional coverage probability of the direct links  multiplied by the BS-TU link probability $\mathrm{Pr}_{\mathrm{LoS}}(\xi)$ defined in \textit{Lemma 7}, and the conditional coverage probability of the reflective LoS links multiplied by the BS-RIS $n$-TU link probability $(1-\mathrm{Pr}_{\mathrm{LoS}}(\xi))P_R^s(\xi)$ defined in \textit{Lemma 8}. Specifically, the conditional SINR probability w.r.t. the direct links is 
\begin{equation}
\label{eq:0046}
\begin{split}
&P_{cov_d|\xi}^{s}(T)=\mathrm{Pr}(\gamma_D>T)\\
&=\Pr\left(\left|h_{d,u,i}\right|^2>\left(\sigma^2+\text{Interference}\right)\frac T{M_rM_t\cdot P\cdot\Omega_{\mathrm{SD}}(\xi)}\right)\\
& = 1- F_{|h_{d,u,i}|^{2}} \left( \left(\sigma^2+\text{Interference}\right)\frac T{M_rM_t\cdot P\cdot\Omega_{\mathrm{SD}}(\xi)}\right)\\
& \; {\approx} 1- F_{h_{d,u,i}} \left( \left(\frac {\sigma^2T}{M_rM_t\cdot P\cdot\Omega_{\mathrm{SD}}(\xi)}\right)^{\frac{1}{2}}\right),
\end{split}
\end{equation}
where $F_{h_{d,u,i}} \left(\cdot\right)$ refers to the CDF expression of the BS-TU channel $|h_{d,u,i}|$ and is given by (\ref{eq:00290}).

For the reflective LoS link selection, we pick out the optimal link $I^{\star}$ providing the largest E2E SINR as the optimal RIS reflective link, according to the selection criterion proposed in (\ref{eq:0015}). Then, the corresponding RIS reflective link SINR yields
\begin{equation}
\label{eq:0047}
\begin{split}
\gamma_{I^{\star}}= &\frac{1}{\sigma^2+\text{Interference}}\left|\sum_{l=1}^{L_{I^{\star}}} h_{n,l} \cdot [\underline{\boldsymbol{\omega}}^{\star}]_{l} \cdot g_{n,l} \right|^2  M_r M_t  P\\
&\cdot \Omega_{\text{S} \text{R{\textit{I*}}}}(\xi) \cdot \Omega_{\text{R{\textit{I*}}D}}(\xi)\\
=& \frac{1}{\sigma^2+\text{Interference}} \left| R_{I^{\star}} \right|^{2} M_{r}M_{t}P\Omega_{\text{S}\text{R{\textit{I*}}}}(\xi) \Omega_{\text{R{\textit{I*}}D}}(\xi).
\end{split}
\end{equation}
Hence, the conditional SINR coverage probability w.r.t. the reflective LoS links can be expressed as
\begin{equation}
\label{eq:0048}
\begin{split}
&P_{cov_I|\xi}^{s}(T) = \mathrm{Pr}(\gamma_{I^{\star}}>T) = \mathrm{Pr} \left(\begin{array}{c}
\dot{r}_{n,l}^{\alpha}\cdot \dot{s}_{n,l}^{\alpha}
<\end{array} \chi_{1} \right)\\
& \approx \int_{0}^{\infty}\int_{0}^{\infty} F_{(\dot{r}_{n,l}\cdot \dot{s}_{n,l})|\xi} (\chi_{2}) f_{|R_{n}|^{2}}(x)\mathrm{d}x,
\end{split}
\end{equation}
where $\chi_{1} = \frac{M_{r}M_{t}\cdot P\cdot \zeta^2 \cdot  \left|R_{n}\right|^{2}}{\left(\sigma^{2}+\mathrm{Interference}\right)T} \approx \frac{M_{r}M_{t}\cdot P\cdot \zeta^2 \cdot  \left|R_{n}\right|^{2}}{\sigma^{2}T}$, and $\chi_{2} = \left(\frac{M_{r}M_{t}\cdot P\cdot\zeta^2\cdot\left|R_{n}\right|^{2}}{(\sigma^{2}+\mathrm{Interference})\cdot T}\right)^{\frac{1}{\alpha}} \approx \left(\frac{M_{r}M_{t}\cdot P\cdot\zeta^2\cdot\left|R_{n}\right|^{2}}{\sigma^{2}\cdot T}\right)^{\frac{1}{\alpha}}$. In addition, the conditional CDF expression $F_{\eta_n|\xi}(x)$ is derived in (\ref{eq:00033}), and the approximate PDF of $R_{n}$ is derived in (\ref{eq:00037}).
Thereby, with a given preset SINR threshold $T$, the conditional coverage probability of users at a distance $\xi$ from BS can be expressed as
\begin{align}
\label{eq:0047}
P_{cov|\xi}^{s}(\mathrm{T}) = \mathrm{Pr}_{\mathrm{Ad}}(\xi) \cdot P_{{cov_{d}|\xi}}^{s}(T)+\mathrm{Pr}_{\mathrm{AI}}(\xi)\cdot P_{{cov_{I}|\xi}}^{s}(T).
\end{align}

Accordingly, the ergodic coverage probability of the cell yields 
\begin{equation}
\small
\label{eq:0048}
\mathbb{E}[P_{cov}^s(T)]
= \frac{\mathbb{E}[\sum_{u}P_{cov|\xi}^s(T) ]}{\int_0^R\lambda_u(\xi)2\pi\xi\mathrm{d}\xi}
\overset{(\text{IV})}{=} \frac{\int_0^R P_{cov|\xi}^s(T)\lambda_u(\xi)2\pi\xi\mathrm{d}\xi}{\int_0^R\lambda_u(\xi)2\pi\xi\mathrm{d}\xi},
\end{equation}
where the step (IV) in formula (\ref{eq:0048}) 
is derived by following the Campbell Theorem \cite{Chiu1989StochasticGA}. This completes the proof of the Theorem 1.
%

%
\end{appendices}

\bibliographystyle{IEEEtran}%
\bibliography{bibfile}

\end{document}